\documentclass[10pt,aps,prd,twocolumn,showpacs,superscriptaddress,eqsecnum,longbibliography,nofootinbib]{revtex4-2}
\usepackage{mathrsfs,amsmath,amsthm,latexsym,amssymb,amsfonts,epsfig,cancel,enumerate,graphicx,txfonts,subfigure}
\usepackage{multirow}
\usepackage{xcolor}
\usepackage[colorlinks,linkcolor=blue,citecolor=blue,urlcolor=blue,plainpages=false,pdfstartview=FitH]{hyperref}
\allowdisplaybreaks
\newcommand{\dd}{\mathrm{d}}
\newcommand{\GZU}{School of Physics, Guizhou University, Guiyang 550025, China}
\usepackage[english]{babel}
\usepackage{lipsum}

\begin{document}

\title{Black hole images as probes of thermodynamic evolution}

\author{Lei You}
\email{gs.lyou25@gzu.edu.cn}
\affiliation{\GZU}

\author{Jinsong Yang}
\thanks{Corresponding author}
\email{jsyang@gzu.edu.cn}
\affiliation{\GZU}

\begin{abstract}
Observable signatures of black hole thermodynamics remain far from fully explored. Previous works have suggested that thermodynamic phase transitions of black holes could leave imprints on their images. In this work, we demonstrate that richer black hole thermodynamic information can also leave imprints on the resulting images. Using the charged anti-de Sitter black hole as an example, we study the evolution of its images (shadow and accretion-disk images) along isobaric and isothermal processes. We find that the image size evolves monotonically along isobars but becomes nonmonotonic along isotherms. After further considering phase transitions, the image size exhibits a sudden increase in both thermodynamic processes. More importantly, in the isothermal process, the phase transition further results in the emergence of a critical reduced temperature that separates two qualitatively distinct image evolutions. These results show that black hole images can probe not only phase transitions, but also thermodynamic process and temperature.
\end{abstract}

\maketitle

\section{Introduction}
\label{sec1}

Black hole thermodynamics has reshaped our understanding of gravity. It reveals that spacetime itself carries an entropy and a temperature and obeys the first law of thermodynamics, suggesting that spacetime admits an underlying statistical description~\cite{Bekenstein:1973ur,Bardeen:1973gs,Hawking:1975vcx}. From this viewpoint, Hawking and Page discovered the phase transition between the thermal gas and the stable large black hole~\cite{Hawking:1982dh}, which was later explained in the anti-de Sitter/conformal field theory (AdS/CFT) correspondence as the confinement-deconfinement phase transition of the dual gauge theory~\cite{Witten:1998zw}. Motivated by this insight, many other phase phenomena were uncovered. Most notably in the extended phase space, where the cosmological constant is treated as a dynamical pressure and the black hole mass as enthalpy~\cite{Kastor:2009wy,Dolan:2010ha,Cvetic:2010jb}, one finds that charged AdS black holes exhibit the Van der Waals-like small/large black hole transition with critical scaling, while various generalizations can display richer structures such as reentrant phase transitions, triple points, and multicritical behavior~\cite{Banerjee:2011au,Banerjee:2011raa,Kubiznak:2012wp,Wei:2012ui,Cai:2013qga,Altamirano:2013uqa,Altamirano:2013ane,Frassino:2014pha,Wei:2014hba,Kubiznak:2016qmn}. These developments show that black hole thermodynamics is not a mere formal analogy but a central bridge linking gravity, quantum theory, and statistical mechanics. This physical significance naturally motivates the search for observable signatures of black hole thermodynamics.

Some early efforts explored quasinormal modes (QNMs) as dynamical probes of black hole thermodynamics, aiming to access thermodynamic information through gravitational-wave observables. In Ref.~\cite{Koutsoumbas:2006xj} the authors found that when a topological AdS black hole undergoes a second-order phase transition to a hairy configuration, the slope of the QNMs spectrum flips sign and the damping time scale changes markedly. This QNMs-transition connection was soon extended to other black holes~\cite{Rao:2007zzb,Koutsoumbas:2008pw,He:2008im}. Notably, Ref.~\cite{Liu:2014gvf} identified a link between QNMs and a Van der Waals-like thermodynamic phase transition in charged AdS spacetime, which helped broaden such studies across a wider class of AdS backgrounds~\cite{Zou:2014sja,Chabab:2017knz,Zou:2017juz,Li:2017kkj}. These results indicate that thermodynamic phase transition information can be encoded in QNMs.

In recent years, horizon-scale imaging has become increasingly mature~\cite{EventHorizonTelescope:2019ggy,EventHorizonTelescope:2022wkp}. This progress has not only spurred growing activity in black hole optics but also attracted increasing attention to the connection between photon geodesics and phase transitions~\cite{Cunha:2019hzj,Bambi:2019tjh,Guerrero:2021ues,Bronzwaer:2021lzo,Vagnozzi:2022moj,Yang:2022btw,Zhang:2023okw,Khodadi:2024ubi,You:2024jeu,You:2024uql}. This attention aims to probe black hole phase transition information through black hole images. In Ref.~\cite{Wei:2017mwc}, the authors found that, at the phase transition in a charged AdS black hole, both the photon sphere radius and the critical impact parameter exhibit a discontinuous jump, indicating a deep link between thermodynamic phase transitions and photon geodesic dynamics. Ref.~\cite{Zhang:2019glo} further examined the behavior of the black hole shadow across the phase transition and also found a discontinuous jump, showing that phase transitions can leave clear imprints on black hole images. Similar conclusions were subsequently confirmed in a variety of spacetimes~\cite{Wei:2018aqm,Zhang:2019tzi,NaveenaKumara:2019nnt,Xu:2019yub,Chabab:2019kfs,Du:2022quq,Ladino:2024ned,Ladino:2025oeq,Yang:2025xck}, and some works further examined the relation between phase transitions and the Lyapunov exponent of particle motion, again finding a discontinuous jump~\cite{Guo:2022kio,Yang:2023hci,Lyu:2023sih,Du:2024uhd,Chen:2025xqc,Xie:2025auj}. Although these studies have been successful in revealing observable signatures of black hole thermodynamics, their limitation is also clear: they mainly access phase transition information. A natural next step is to explore whether other thermodynamic information of black holes can also be observed.

Very recently, Ref.~\cite{You:2025apm} provided a possible clue: during isobaric and isothermal thermodynamic processes of the charged AdS black hole, the photon sphere radius $r_{\rm ps}$ varies monotonically and nonmonotonically with the horizon radius $r_h$, respectively. This difference leads us to the following considerations. According to the black hole area law, the horizon area~\cite{Bardeen:1973gs}, or equivalently the horizon radius, should increase with time, i.e., $r_h(t)$ is monotonically increasing. Although Hawking radiation provides a mechanism that can shrink the horizon~\cite{Hawking:1975vcx}, this effect is negligible for astrophysical black holes. It is therefore reasonable to assume ${\rm d}r_h/{\rm d}t>0$. By the chain rule,
\begin{equation}
 \frac{{\rm d}r_{\rm ps}}{{\rm d}t}=\frac{{\rm d}r_{\rm ps}}{{\rm d}r_h}\,\frac{{\rm d}r_h}{{\rm d}t},
\end{equation}
thus ${\rm d}r_{\rm ps}/{\rm d}t$ and ${\rm d}r_{\rm ps}/{\rm d}r_h$ share the same sign. Therefore, the monotonic or nonmonotonic behavior of $r_{\rm ps}(r_h)$ is inherited by $r_{\rm ps}\big[r_h(t)\big]$. This allows one to view the evolution of $r_{\rm ps}$ in black hole thermodynamic processes from a time-evolution perspective. More importantly, the evolution pattern of $r_{\rm ps}$ is likely to be inherited by black hole images. If so, black holes evolving isobarically and isothermally would exhibit distinct imaging evolutions. This distinction is likely to be discovered in long-term astronomical observations, and thus allow one to extract information about the black hole's thermodynamic process. Moreover, in this work we will show that this time-evolution perspective can also extract temperature information from black hole images. Briefly, earlier studies were largely restricted to monotonic $r_{\rm ps}(r_h)$ and could therefore extract only phase transition signatures. By extending the analysis to nonmonotonic behavior, we aim to broaden the avenues for extracting black hole thermodynamic information from images. Finally, we would like to emphasize that we do not claim astrophysical black holes strictly follow idealized thermodynamic processes. Rather, our goal is to use such idealized settings to identify what kinds of thermodynamic information may, in principle, be accessible, which can still offer useful guidance for realistic observations.

To implement the above idea, we take the Reissner-Nordstr\"om--AdS (RN--AdS) black hole as a concrete example and investigate how its shadow and accretion-disk images evolve along isobaric and isothermal thermodynamic processes. The paper is organized as follows. In Sect.~\ref{sec2}, we briefly review the basic framework of black hole shadow and accretion-disk imaging to set the stage for later analysis. In Sect.~\ref{sec3}, combining analytic arguments with numerical calculations, we study the evolution of these images in the two thermodynamic processes, relate their behaviors to black hole phase transitions, and explore how additional thermodynamic information can be extracted from black hole images. Finally, Sect.~\ref{sec4} contains our discussion and conclusions. Throughout this paper we work in natural units with $G=c=\hbar=k_B=1$.

\section{Geodesics and Black Hole Imaging}
\label{sec2}

In this section we briefly review the key formulas of black hole imaging for use in the next section. The line element of the four-dimensional RN--AdS black hole spacetime reads
\begin{equation}
 ds^{2}=-f(r)\, \dd t^{2} + f(r)^{-1} \dd r^{2} + r^{2}\!\left(\dd\theta^{2} + \sin^{2}\theta\, \dd\phi^{2}\right),
 \label{eq:metric1}
\end{equation}
with
\begin{equation}
 f(r) = 1 - \frac{2M}{r} + \frac{Q^{2}}{r^{2}} + \frac{r^{2}}{\ell^{2}}.
 \label{eq:metric2}
\end{equation}
Here $M$ and $Q$ denote the black hole mass and electric charge, respectively, while $\ell$ is the AdS curvature radius associated with the negative cosmological constant. In the extended phase space formulation of black hole thermodynamics, $\ell$ is related to the thermodynamic pressure $P$ via
\begin{equation}
 \ell = \sqrt{\frac{3}{8\pi P}},
 \qquad
 P = \frac{3}{8\pi \ell^{2}}.
\end{equation}
In the limit $\ell\to+\infty$ (equivalently $P\to 0^+$), the metric~\eqref{eq:metric1} reduces to the asymptotically flat RN spacetime.

The motion of a test particle in a static, spherically symmetric spacetime is described by the Lagrangian
\begin{equation}
 2\mathcal{L} = g_{\mu\nu}\dot x^{\mu}\dot x^{\nu} = -\varepsilon,
 \quad
 \varepsilon =
 \begin{cases}
 0, & \text{massless particle (photon)},\\[2pt]
 1, & \text{massive particle},
 \end{cases}
 \label{eq:l1}
\end{equation}
where the overdot denotes differentiation with respect to an affine parameter. Substituting the metric~\eqref{eq:metric1} into the above expression yields
\begin{equation}
 2\mathcal{L}= - f(r)\, \dot t^{2} + f(r)^{-1} \dot r^{2} + r^{2}\dot\theta^{2} + r^{2}\sin^{2}\theta\, \dot\phi^{2}=-\varepsilon,
 \label{eq:l2}
\end{equation}
which does not depend explicitly on $t$ or $\phi$, and hence allows us to define two conserved quantities,
\begin{align}
 E &:= -\frac{\partial\mathcal{L}}{\partial \dot t} = f(r)\,\dot t,\\
 L &:= \frac{\partial\mathcal{L}}{\partial \dot\phi} = r^{2}\sin^{2}\theta\,\dot\phi,
 \label{eq:phidot}
\end{align}
corresponding to the energy and angular momentum of the particle, respectively. Solving the above relations for $\dot t$ and $\dot\phi$ and substituting them into Eq.~\eqref{eq:l2}, one obtains the radial equation of motion (fixing $\theta=\pi/2$),
\begin{equation}
 \dot r^{2} = E^{2} - f(r)\!\left(\varepsilon + \frac{L^{2}}{r^{2}}\right).
 \label{eq:rdot1}
\end{equation}
For photons $(\varepsilon=0)$, the energy $E$ can be absorbed by the reparametrization $\lambda\!\rightarrow\!\lambda E$, and the radial equation then simplifies to
\begin{equation}
 \dot r^{2} = 1 - b^{2}\,U(r),
 \label{eq:rdot2}
\end{equation}
where $b:=L/E$ is the impact parameter and $U(r):=f(r)/r^{2}$ is the effective potential.
\begin{figure}[htbp]
 \centering
 \includegraphics[width=1\linewidth]{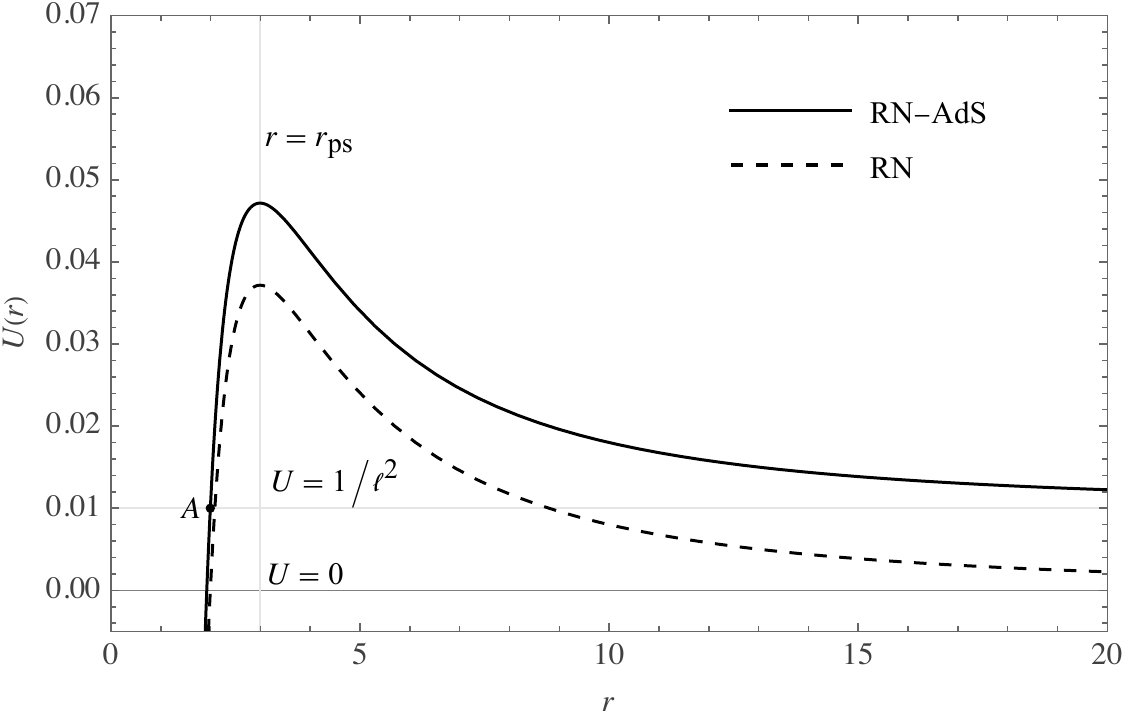}

 \caption{Effective potential $U(r)$ for the RN--AdS and asymptotically flat RN black holes. The horizontal axis $r$ has length dimension $L$, while the vertical axis $U$ has dimension $L^{-2}$.}
 \label{fig:ur}
\end{figure}

\begin{figure}[htbp]
 \centering
 \includegraphics[width=1\linewidth]{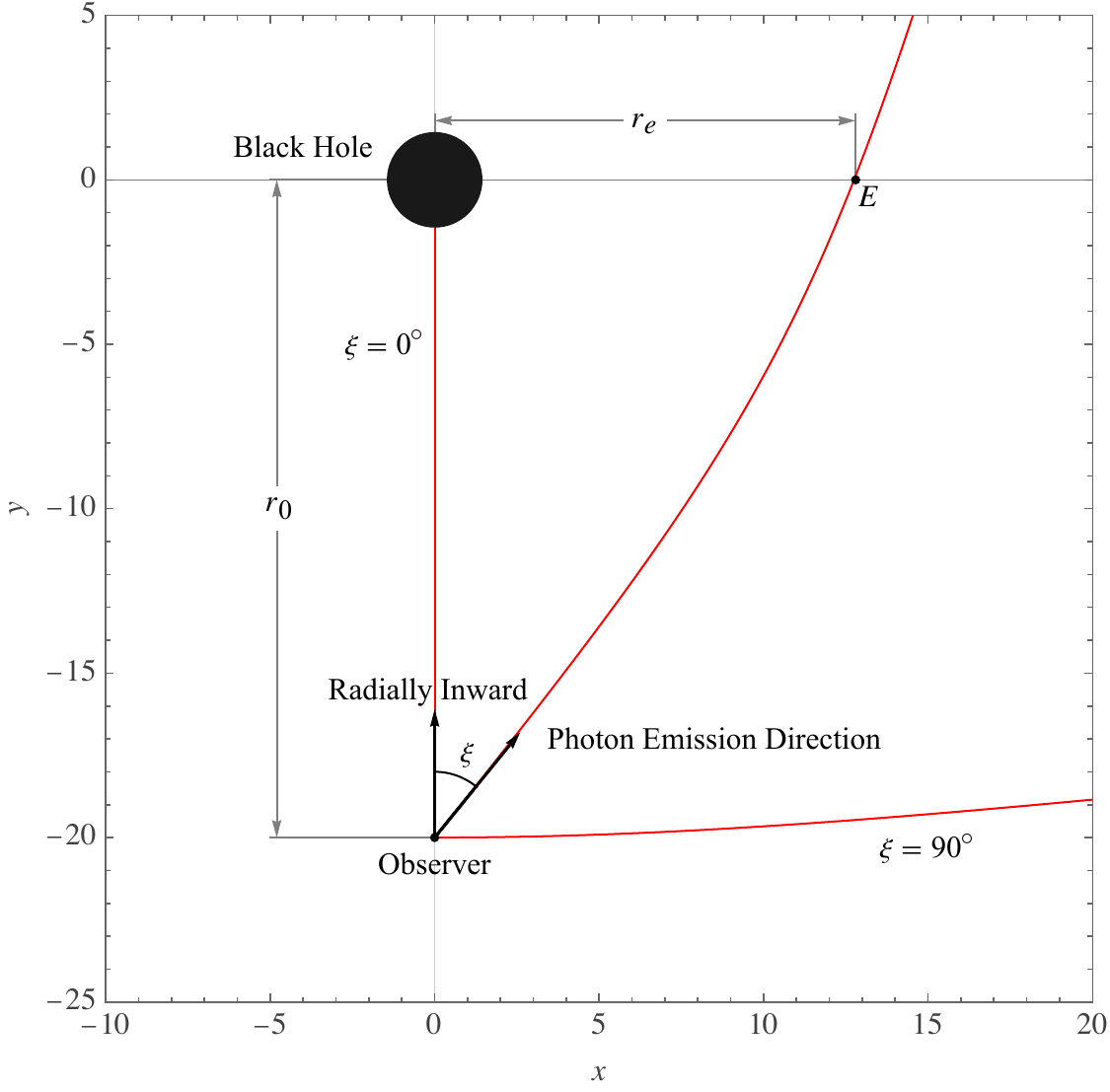}

 \caption{Schematic illustration of the ray-tracing method. The observer is located at a distance $r_0$ from the black hole, and point $E$ denotes an accretion-disk particle at a distance $r_e$ from the black hole. The horizontal axis $x$, the vertical axis $y$, and the quantities $r_0$ and $r_e$ all have length dimension $L$, whereas $\xi$ is dimensionless.}
 \label{fig:ob}
\end{figure}

Figure~\ref{fig:ur} compares the effective potentials of RN--AdS and RN black holes, showing that the photon dynamics is qualitatively similar in the two spacetimes. However, a significant difference arises at large radii: the RN effective potential vanishes asymptotically, while the RN--AdS potential approaches a positive constant,
\begin{equation}
 \lim_{r\to+\infty} U(r) = \lim_{r\to+\infty} \left( \frac{1}{r^{2}} - \frac{2M}{r^{3}} + \frac{Q^{2}}{r^{4}} + \frac{1}{\ell^{2}} \right)
 = \frac{1}{\ell^{2}}.
\end{equation}
As a consequence, the condition $\dot r^{2}\ge0$ implies that photons with $b\ge\ell$ can propagate only within a finite radial interval $r\le r_{A}$, where $r_{A}$ is defined by the intersection of $U(r)$ with $1/\ell^{2}$ (point~$A$ in Fig.~\ref{fig:ur}). Therefore, for an observer located at $r_0>r_{A}$, such photons can never reach the observer and do not contribute to the black hole image. In this work the observer is always placed outside the unstable photon sphere, i.e., at $r_0>r_{\rm ps}$, where $r_{\rm ps}$ is defined by $\left.{\rm d}U/{\rm d}r\right|_{r=r_{\rm ps}}=0$. Since $r_{\rm ps}$ always lies outside $r_{A}$, the condition $r_{0}>r_{A}$ is automatically satisfied. Finally, only photons with $b<\ell$ contribute to the image, and their trajectories on the equatorial plane obtained by eliminating the affine parameter between Eqs.~\eqref{eq:phidot} and \eqref{eq:rdot2} obey
\begin{equation}
 \frac{{\rm d}r}{{\rm d}\phi} = \pm\,\frac{r^{2}}{b}\, \sqrt{\,1 - b^{2}\,\frac{f(r)}{r^{2}}\,},
\end{equation}
where “$+$” (“$-$”) corresponds to outgoing (ingoing) motion. We denote its solution by $r(\phi\,;b)$, which allows us to construct the accretion-disk image in a ray-tracing scheme. Figure~\ref{fig:ob} illustrates the schematic of the ray-tracing method. A photon is emitted from the observer at an angle $\xi$ relative to the inward radial direction toward the black hole, reaching the accretion disk at point $E$ with radius $r_e$. Due to the reversibility of light paths, this trajectory is equivalent to a photon emitted from point $E$ on the accretion disk reaching the observer. Therefore, the angle $\xi$ can be directly interpreted as the apparent angular size of the image formed by point $E$.

The angle $\xi$ is related to the photon impact parameter $b$ by~\cite{Perlick:2021aok} (see Appendix~\ref{appa} for the derivation)
\begin{equation}
 \sin\xi = \frac{b\,\sqrt{f(r_0)}}{r_0}.
 \label{eq:sinxi}
\end{equation}
All photons that can reach the accretion disk satisfy
\begin{equation}
 r\!\left(\phi_n\,; b_n(\xi_n)\right) = r_e,
 \label{eq:pfun}
\end{equation}
where $r_e$ denotes the orbital radius of the emitting disk particle. In this work we take
\begin{equation}
 r_e \in [\,r_{\rm isco},\,+\infty),
\end{equation}
with $r_{\rm isco}$ being the radius of the innermost stable circular orbit (ISCO) of massive particles $(\varepsilon=1)$, obtained by solving the following three conditions:
\begin{equation}
 \dot r (r;E,L) = 0, \quad
 \frac{{\rm d}\dot r(r;E,L)}{{\rm d}r} = 0, \quad
 \frac{{\rm d}^{2}\dot r(r;E,L)}{{\rm d}r^{2}} = 0,
\end{equation}
where $\dot r(r;E,L)$ is given by the radial equation~\eqref{eq:rdot1} evaluated at $\varepsilon=1$. The azimuthal angle $\phi_n$ denotes the angle accumulated by a photon when it intersects the accretion disk. Since we consider only observers located above the north pole of the black hole, $\phi_n$ takes the simple form
\begin{equation}
 \phi_n = -\frac{\pi}{2} + n\,\pi,
 \qquad n=1,2,\dots
\end{equation}
which correspond to the standard primary ($n=1$) and secondary ($n=2$) images in classical accretion-disk lensing theory. Then, solving Eq.~\eqref{eq:pfun} for $b_n$ and substituting it into Eq.~\eqref{eq:sinxi}, the angular size of the accretion-disk image is obtained as
\begin{equation}
 \xi_n = \arcsin\!\left(\frac{b_n\,\sqrt{f(r_0)}}{r_0}\right),
 \label{eq:xin}
\end{equation}
which in general admits no closed-form expression and is therefore evaluated numerically in the next section. By contrast, the angular size of the black hole shadow can be obtained analytically as
\begin{equation}
 \xi_{\rm c} = \arcsin\!\left(\frac{b_{\rm c}\,\sqrt{f(r_0)}}{r_0}\right),
 \label{eq:xic}
\end{equation}
where
\begin{equation}
 b_{\rm c} = \frac{r_{\rm ps}}{\sqrt{f(r_{\rm ps})}}
\end{equation}
is the critical impact parameter separating photons captured by the black hole from those escaping to large distances. In the next section, we will use Eqs.~\eqref{eq:xin} and \eqref{eq:xic} to analyze how the black hole image, including both the black hole shadow and the accretion-disk image, evolves along different thermodynamic processes. It should be emphasized that a complete analysis of the accretion-disk image should include an emission model and radiative-transfer effects, such as emissivity, absorption, scattering, and frequency-dependent intensity modulation. In this paper, however, we focus only on the angular size of the accretion disk, which is sufficient for our purpose. Therefore, our subsequent results reflect only the geometric response of the accretion disk to thermodynamic evolution, rather than its full physical response.

\section{Image Evolution in Different thermodynamic processes}
\label{sec3}

In this section we investigate how black hole images evolve along isobaric and isothermal processes. Before proceeding, we briefly recall the distinct behavior of the photon sphere radius $r_{\rm ps}$ in these two thermodynamic processes, which is governed by the monotonicity criterion~\cite{You:2025apm}
\begin{equation}
 \frac{{\rm d} r_{\rm ps}}{{\rm d} r_h} = -\,\frac{1}{\Gamma_0}\, \frac{{\rm d} M}{{\rm d} r_h}\, \partial_M\Phi_0.
 \label{eq:pj}
\end{equation}
For the RN--AdS black hole, $\Gamma_0<0$ while $\partial_M\Phi_0>0$, and thus, along any thermodynamic process, the monotonicity of $r_{\rm ps}(r_h)$ is determined solely by ${\rm d}M/{\rm d}r_h$ (see Ref.~\cite{You:2025apm} for a detailed analysis). Since ${\rm d}M/{\rm d}r_h$ behaves differently in isobaric and isothermal processes, the photon sphere radius exhibits qualitatively distinct behavior in the two processes.

\begin{figure}[htbp]
 \centering
 \includegraphics[width=1\linewidth]{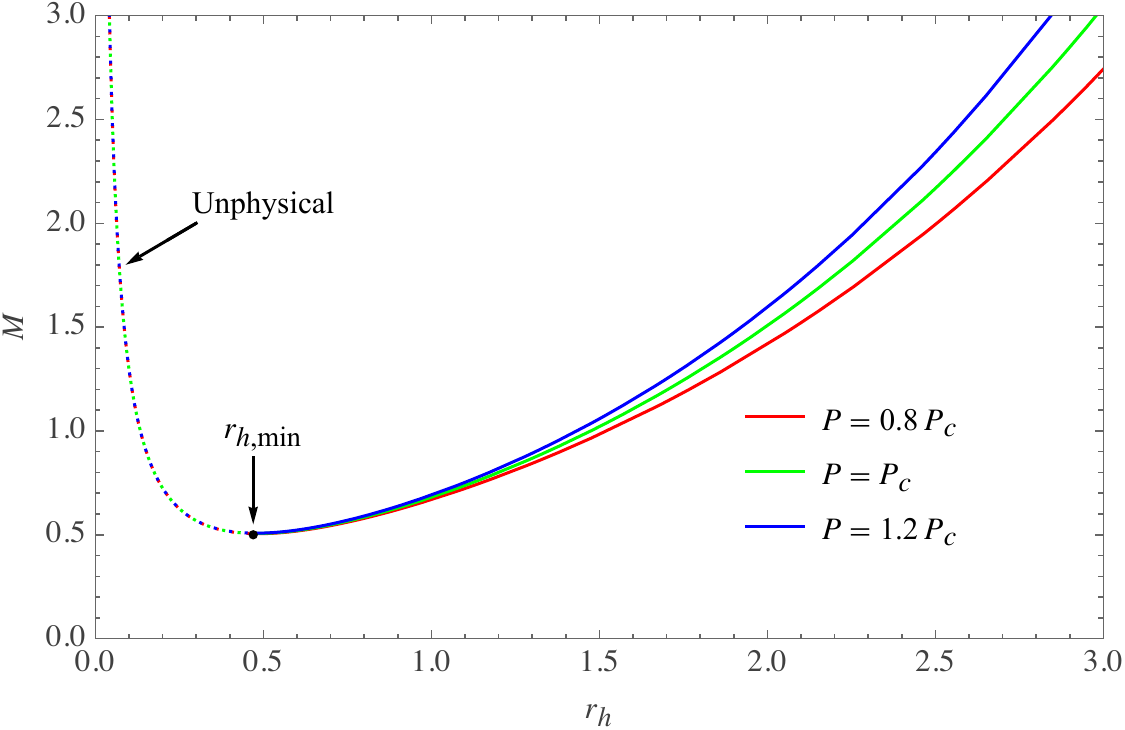}

 \caption{Isobaric $M(r_h)$ curves for $Q=0.5$. The horizontal axis $r_h$, the vertical axis $M$, as well as the quantities $r_{h,\rm min}$ and the fixed parameter $Q$, all have length dimension $L$, whereas $P$ and $P_c$ have dimension $L^{-2}$.}
 \label{fig:xdy}
\end{figure}

In the isobaric process, the pressure $P$ is fixed and $M(r_{h})$ is determined from $f(r_{h};M)=0$, yielding
\begin{equation}
 M(r_{h}) = \frac{3 Q^{2} + 3 r_{h}^{2} + 8\pi P\, r_{h}^{4}}{6 r_{h}}.
 \label{eq:Mrh_isobaric}
\end{equation}
Mathematically this function decreases for small $r_{h}$ before increasing at larger $r_{h}$. Physically, however, the first law of black hole thermodynamics, ${\rm d}M = T\,{\rm d}S$, imposes the constraint
\begin{equation}
 \frac{{\rm d}M}{{\rm d}r_{h}} = 2\pi r_{h} T > 0,
 \label{eq:ddmrh1}
\end{equation}
where $T=\partial_r f(r_h)/4\pi$ denotes the Hawking temperature and $S=\pi r_{h}^{2}$ is the Bekenstein-Hawking entropy. Consequently only the monotonically increasing branch is admissible, and the curve must be truncated at its minimum,
\begin{equation}
 r_{h,\min} = \frac{1}{4} \sqrt{-\frac{1}{\pi P} + \frac{\sqrt{\,1 + 32\pi P\, Q^{2}\,}}{\pi P}}.
\end{equation}
Figure~\ref{fig:xdy} displays the $M(r_h)$ curves for pressures above, below, and equal to the critical pressure $P_c$, where $P_c$ is obtained by solving the following two criticality conditions:
\begin{equation}
 \frac{\partial T(r_h;P)}{\partial r_h}=0,
 \qquad
 \frac{\partial^{2} T(r_h;P)}{\partial r_h^{2}}=0.
\end{equation}
For the RN--AdS black hole, the solutions to the above equations are
\begin{equation}
 r_{h,c}=\sqrt{6}\,Q, \qquad P_c=\frac{1}{96\pi\,Q^{2}}.
 \label{eq:rhp}
\end{equation}

\begin{figure}[htbp]
 \centering
 \includegraphics[width=1\linewidth]{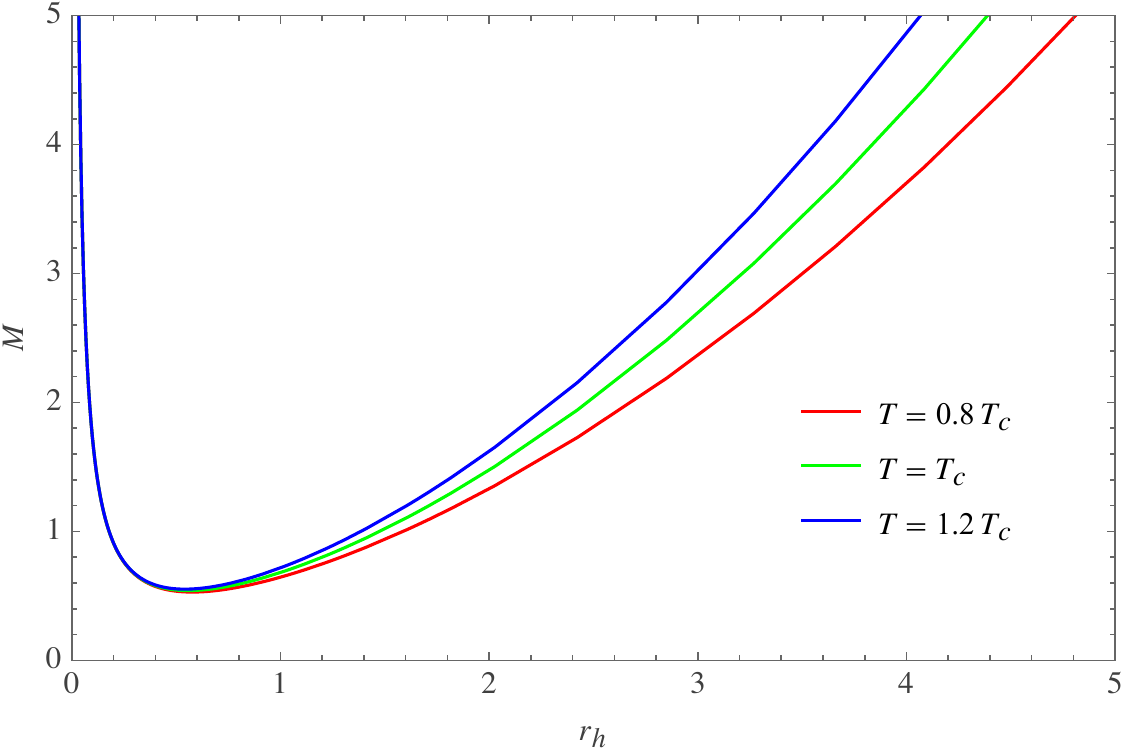}

 \caption{Isothermal $M(r_h)$ curves for $Q=0.5$. The horizontal axis $r_h$, the vertical axis $M$, and the fixed parameter $Q$ all have length dimension $L$, whereas $T$ and $T_c$ have dimension $L^{-1}$.}
 \label{fig:xdt2}
\end{figure}

\begin{figure}[htbp]
 \centering
 \includegraphics[width=1\linewidth]{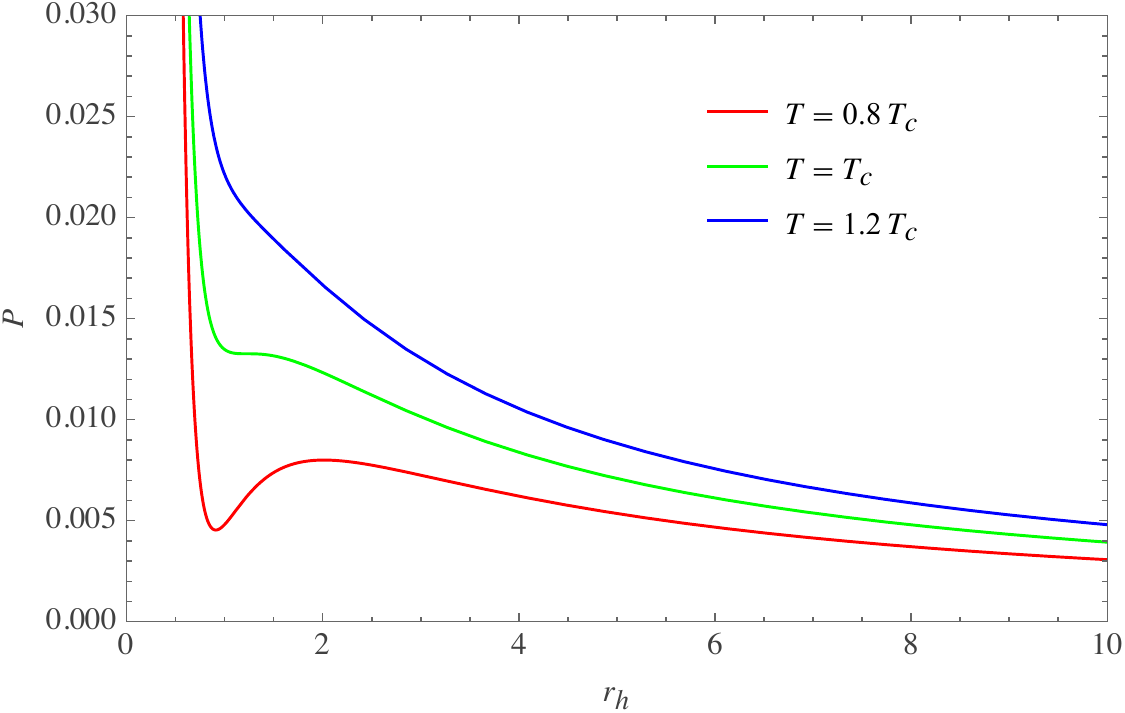}

 \caption{Isothermal $P(r_h)$ curves for $Q=0.5$. The horizontal axis $r_h$ has length dimension $L$, while the vertical axis $P$ has dimension $L^{-2}$, and the quantities $T$ and $T_c$ have dimension $L^{-1}$.}
 \label{fig:xdt1}
\end{figure}

In contrast, along the isothermal process, both $M$ and $P$ vary and are constrained by the horizon condition $f(r_{h};M,P)=0$ together with the isothermality condition $T(r_{h};M,P)=T_{0}$, which yield
\begin{align}
 M(r_{h}) &= \frac{2 Q^{2} + r_{h}^{2} + 2\pi T_{0}\, r_{h}^{3}}{3 r_{h}},
 \label{eq:mrh} \\
 P(r_{h}) &= \frac{Q^{2} - r_{h}^{2} + 4\pi T_{0}\, r_{h}^{3}}{8\pi r_{h}^{4}}.
 \label{eq:prh}
\end{align}
Mathematically, $M(r_h)$ again decreases at small $r_h$ before increasing at larger $r_h$. Physically, however, the first law now takes its extended form, ${\rm d}M = T\,{\rm d}S + V\,{\rm d}P$, which implies
\begin{equation}
 \frac{{\rm d}M}{{\rm d}r_h} = 2\pi r_h\,T + V\,\frac{{\rm d}P}{{\rm d}r_h}.
 \label{eq:ddmrh2}
\end{equation}
Compared with Eq.~\eqref{eq:ddmrh1}, the extended first law introduces an additional term $V\,{\rm d}P/{\rm d}r_h$. The thermodynamic volume satisfies $V>0$ and ${\rm d}P/{\rm d}r_h$ exhibits a characteristic van der Waals–type behavior, which becomes strongly negative at small $r_h$. As a result, ${\rm d}M/{\rm d}r_h$ can vanish or even turn negative, allowing $M(r_h)$ to avoid truncation and retain its nonmonotonicity. Figures~\ref{fig:xdt2} and \ref{fig:xdt1} display the $M(r_h)$ and $P(r_h)$ curves for temperatures above, below, and equal to the critical value $T_c$, respectively, where $T_c$ is likewise obtained by solving the two criticality conditions:
\begin{equation}
 \frac{\partial P(r_h;T)}{\partial r_h}=0,
 \qquad
 \frac{\partial^2 P(r_h;T)}{\partial r_h^2}=0\,.
\end{equation}
For the RN--AdS black hole, the above equations give
\begin{equation}
 r_{h,c}=\sqrt{6}\,Q, \qquad
 T_{c} = \frac{1}{3\sqrt{6}\,\pi\,Q}.
 \label{eq:rht}
\end{equation}

\begin{figure}[htbp]
 \centering
 \includegraphics[width=1\linewidth]{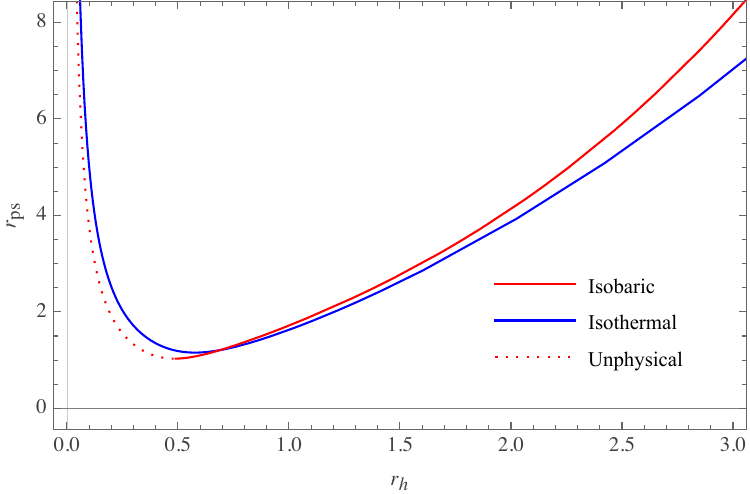}

 \caption{Photon sphere radius $r_{\rm ps}$ as a function of the horizon radius $r_h$ for isobaric ($P=0.8P_c$) and isothermal ($T=0.8T_c$) processes. For both the isobaric and isothermal cases, we set $Q=0.5$. The horizontal axis $r_h$, the vertical axis $r_{\rm ps}$, and the fixed charge $Q$ all have length dimension $L$, while $P$ and $T$ have dimensions $L^{-2}$ and $L^{-1}$, respectively.}
 \label{fig:rps}
\end{figure}

These distinct behaviors of $M(r_h)$ in the two thermodynamic processes lead directly to different monotonicity of $r_{\rm ps}(r_h)$. Figure~\ref{fig:rps} displays the corresponding numerical results, which are fully consistent with the above analysis. Given the central role of the photon sphere in black hole imaging, it is then natural to expect that both the shadow angle $\xi_{\rm c}(r_h)$ and the accretion-disk image angle $\xi_n(r_h)$ inherit these monotonic or nonmonotonic features. More importantly, these features can in principle be revealed by long-term astronomical observations. Specifically, the actual evolution of a black hole may involve many complicated nonequilibrium effects, such as nonstationary accretion, radiative processes, and nonequilibrium heat exchange. However, in the framework of black hole thermodynamics, we do not attempt to model these detailed dynamical processes directly, but instead consider idealized quasi-static processes in which the black hole passes through a sequence of nearby equilibrium states. Under this assumption, the state of the black hole can be characterized by thermodynamic variables. For the RN--AdS black hole studied here, the equilibrium states can be characterized by variables such as the mass $M$, pressure $P$, temperature $T$, and horizon radius $r_h$. These variables naturally define a thermodynamic state space, and the evolution process of the black hole can then be represented by a path in this space. In our analysis, both the isobaric and isothermal cases are treated as such idealized quasi-static paths, and both can be conveniently parameterized by the horizon radius $r_h$. This leads to the corresponding relations $M(r_h)$, $P(r_h)$, and $T(r_h)$ along the chosen path. The shadow angle $\xi_{\rm c}$ and the accretion-disk image angle $\xi_n$ are determined by the black hole state and can therefore be parameterized by $r_h$. Similarly, we can parameterize the evolution path in the state space by time $t$, which physically corresponds to the growth of the black hole. Since the black hole area law implies that $r_h(t)$ is monotonic, the qualitative behaviors of $\xi_{\rm c}(r_h)$ and $\xi_n(r_h)$ are the same as those of $\xi_{\rm c}[r_h(t)]$ and $\xi_n[r_h(t)]$. In this sense, $r_h$ can be naturally regarded as an effective proxy for black hole growth. Therefore, the behavior of $\xi_{\rm c}(r_h)$ and $\xi_n(r_h)$ allows us to infer what may be observed in black hole images over long-term astronomical observations and to extract the corresponding thermodynamic information from such image behavior. It should be emphasized, however, that the isobaric and isothermal processes considered here are idealized quasi-static processes, which assume that the black hole evolves respectively at fixed pressure and fixed temperature. In realistic black hole evolution, such idealized conditions may not be strictly satisfied, but this does not prevent us from using them to analyze the basic thermodynamic behavior of black holes. In fact, in ordinary thermodynamics, employing idealized quasi-static processes to study the essential thermodynamic behavior of real fluids is a standard practice.

We now begin a systematic analysis of how black hole images evolve under different thermodynamic processes, combining analytic arguments with numerical calculations. The analytic part focuses on the black hole shadow, since only the critical angle $\xi_{\rm c}$ admits a simple closed-form expression, whereas accretion-disk images generally do not.

The shadow angle given in Eq.~\eqref{eq:xic} can be equivalently written as
\begin{equation}
 \xi_{\rm c} =\arcsin\!\left[ \frac{\sqrt{U(r_0)}}{\sqrt{U(r_{\rm ps})}} \right]
 \equiv \arcsin\,(h).
 \label{eq:ubu}
\end{equation}
where $h$ (and hence $\xi_{\rm c}$) depends on $r_h$ through $M(r_h)$, $\ell(r_h)$ [equivalently $P(r_h)$], and $r_{\rm ps}(r_h)$. Since the $\arcsin$ function is monotonically increasing on its domain, $\xi_{\rm c}(r_h)$ and $h(r_h)$ have the same monotonicity properties, and it therefore suffices to analyze the latter. Differentiating $h(r_h)$ with respect to $r_h$ yields
\begin{equation}
 \frac{{\rm d}h}{{\rm d}r_h} = \frac{g}{2\,U(r_{\rm ps})},
\end{equation}
with
\begin{equation}
 g= h^{-1}\,\frac{{\rm d}U(r_0)}{{\rm d}r_h} - h\,\frac{{\rm d}U(r_{\rm ps})}{{\rm d}r_h}.
\end{equation}
Since $U(r_{\rm ps})>0$, the monotonicity of $h(r_h)$ (and hence of $\xi_{\rm c}(r_h)$) is entirely determined by the sign of $g$. For convenience, we instead examine
\begin{equation}
 h\,g = \frac{{\rm d}U(r_0)}{{\rm d}r_h} - h^{2}\,\frac{{\rm d}U(r_{\rm ps})}{{\rm d}r_h},
 \label{eq:hg1}
\end{equation}
whose sign is identical to that of $g$ due to $0< h\le1$. Using the chain rule (with $P$ replacing $\ell$), one obtains
\begin{equation}
 \frac{{\rm d}U(r_0)}{{\rm d}r_h} = \frac{\partial U(r_0)}{\partial M}\, \frac{{\rm d}M}{{\rm d}r_h} + \frac{\partial U(r_0)}{\partial P}\, \frac{{\rm d}P}{{\rm d}r_h},
\end{equation}
and
\begin{equation}
 \frac{{\rm d}U(r_{\rm ps})}{{\rm d}r_h}
 =
 \frac{\partial U(r_{\rm ps})}{\partial M}\,
 \frac{{\rm d}M}{{\rm d}r_h}
 +
 \frac{\partial U(r_{\rm ps})}{\partial P}\,
 \frac{{\rm d}P}{{\rm d}r_h}
 +
 \frac{\partial U(r_{\rm ps})}{\partial r_{\rm ps}}\,
 \frac{{\rm d}r_{\rm ps}}{{\rm d}r_h},
\end{equation}
where the last term ${\partial U(r_{\rm ps})}/{\partial r_{\rm ps}}$ vanishes identically since $r_{\rm ps}$ satisfies the photon sphere condition
\begin{equation}
 \frac{\partial U(r)}{\partial r}\bigg|_{r=r_{\rm ps}}=0.
\end{equation}
The remaining partial derivatives for the RN--AdS spacetime are
\begin{equation}
 \frac{\partial U(r_0)}{\partial M}=-\frac{2}{r_{0}^{3}},
 \qquad
 \frac{\partial U(r_{\rm ps})}{\partial M}=-\frac{2}{r_{\rm ps}^{3}},
\end{equation}
and
\begin{equation}
 \frac{\partial U(r_0)}{\partial P}=\frac{\partial U(r_{\rm ps})}{\partial P}=\frac{8\pi}{3}.
\end{equation}
Substituting these results into Eq.~\eqref{eq:hg1} yields
\begin{equation}
 h\,g=k_1\,\frac{{\rm d}M}{{\rm d}r_h}+ k_2\,\frac{{\rm d}P}{{\rm d}r_h},
 \label{eq:hg2}
\end{equation}
with
\begin{equation}
 k_1=\frac{2h^{2}}{r_{\rm ps}^{3}}-\frac{2}{r_{0}^{3}},\qquad
 k_2=\frac{8\pi\,(1-h^{2})}{3}.
\end{equation}
Both coefficients satisfy $k_1,k_2\ge0$, with equality attained only when $r_0=r_{\rm ps}$. The nonnegativity of $k_2$ follows directly from $0< h\le1$, while $k_1$ can be rewritten as
\begin{equation}
 k_1=\frac{2}{r_{0}^{2}}\left[\frac{f(r_0)}{f(r_{\rm ps})}\,\frac{1}{r_{\rm ps}}-\frac{1}{r_0}\right],
\end{equation}
which is manifestly nonnegative for $r_0\ge r_{\rm ps}$. These results demonstrate that the monotonicity of $\xi_{\rm c}(r_h)$ is jointly controlled by ${\rm d}M/{\rm d}r_h$ and ${\rm d}P/{\rm d}r_h$.

In the isobaric process the pressure is fixed, ${\rm d}P/{\rm d}r_h=0$, and the monotonicity of $\xi_{\rm c}(r_h)$ is therefore entirely governed by ${\rm d}M/{\rm d}r_h$. As discussed earlier, the first law enforces ${\rm d}M/{\rm d}r_h>0$ on the physical branch, which directly implies that $\xi_{\rm c}(r_h)$ is monotonic for all $r_h>r_{h,{\rm min}}$. Moreover, since $r_{\rm ps}(r_h)$ increases monotonically with $r_h$ (see Fig.~\ref{fig:rps}), it approaches the observer position $r_0$ as $r_h$ grows. As a result, the observer becomes effectively closer to the black hole in an optical sense, driving the shadow angle toward $\xi_{\rm c}\to 90^\circ$.

By contrast, in the isothermal process the pressure is no longer fixed but exhibits a characteristic van der Waals-like behavior, so that both ${\rm d}M/{\rm d}r_h$ and ${\rm d}P/{\rm d}r_h$ contribute to $hg$. Then, from Eq.~\eqref{eq:hg2} one sees that when ${\rm d}M/{\rm d}r_h=0$ the sign of $hg$ is determined by ${\rm d}P/{\rm d}r_h$, and the first law in Eq.~\eqref{eq:ddmrh2} implies ${\rm d}P/{\rm d}r_h<0$ at this point. Similarly, when ${\rm d}P/{\rm d}r_h=0$ the sign of $hg$ is fixed by ${\rm d}M/{\rm d}r_h$, for which Eq.~\eqref{eq:ddmrh2} gives ${\rm d}M/{\rm d}r_h>0$. In other words, $hg<0$ at the extremum of $M(r_h)$ and $hg>0$ at the extremum of $P(r_h)$. It follows that $hg$ must change sign between these two extrema, so $\xi_{\rm c}(r_h)$ is globally nonmonotonic, with its extremum lying between the extrema of $M(r_h)$ and $P(r_h)$. It is worth noting that for $T\le T_c$ the curve $P(r_h)$ has extrema, whereas for $T>T_c$ it has no extrema but still approaches $P\to 0$ as $r_h\to+\infty$ (see Fig.~\ref{fig:xdt1}). Therefore, the above analysis applies to all temperatures. Moreover, as in the isobaric process, for sufficiently small or sufficiently large $r_h$ one finds $r_{\rm ps}\!\to\!r_0$ (see Fig.~\ref{fig:rps}), so that $\xi_{\rm c}\!\to\!90^\circ$.

Before presenting the numerical results, we would like to emphasize that the above analysis is consistent with the expectations outlined in Sect.~\ref{sec1}. In the two thermodynamic processes, the black hole shadow angle indeed exhibits distinct monotonicity properties, which can in principle serve as a criterion for distinguishing the thermodynamic processes in astronomical observations. It is worth noting that this conclusion is qualitatively independent of the observer position $r_0$. The reason is simple: the monotonicity of $\xi_{\rm c}(r_h)$ is completely determined by ${\dd}M/{\dd}r_h$ and ${\dd}P/{\dd}r_h$, both of which are independent of $r_0$. The effect of $r_0$ on $\xi_{\rm c}$ is mainly quantitative: a larger $r_0$ leads to a smaller value of $\xi_{\rm c}$. In other words, an observer farther from the black hole sees a smaller black hole image. This can be readily seen from Eq.~\eqref{eq:ubu} together with Fig.~\ref{fig:ur}. In addition, the $U(r)$ curve in Fig.~\ref{fig:ur} shows that when $r_0$ is large, $U(r_0)$ becomes flat. This indicates that when the observer is sufficiently far from the black hole, the shadow angle $\xi_{\rm c}$ is not very sensitive to changes in the observer position. By contrast, when the observer is close to the photon sphere, $\xi_{\rm c}$ becomes relatively more sensitive to the observer position. In the numerical calculations below, we set $r_0=50$ without loss of generality. Moreover, we note that the distinct monotonicity properties in the two thermodynamic processes are closely tied to the first law of black hole thermodynamics. Given the universality of this law, we expect similar behavior to arise in other black hole models, provided that $k_1,k_2\ge0$ and that $M(r_h)$ and $P(r_h)$ each admit an extremum.

\begin{figure}[htbp]
 \centering
 \includegraphics[width=1\linewidth]{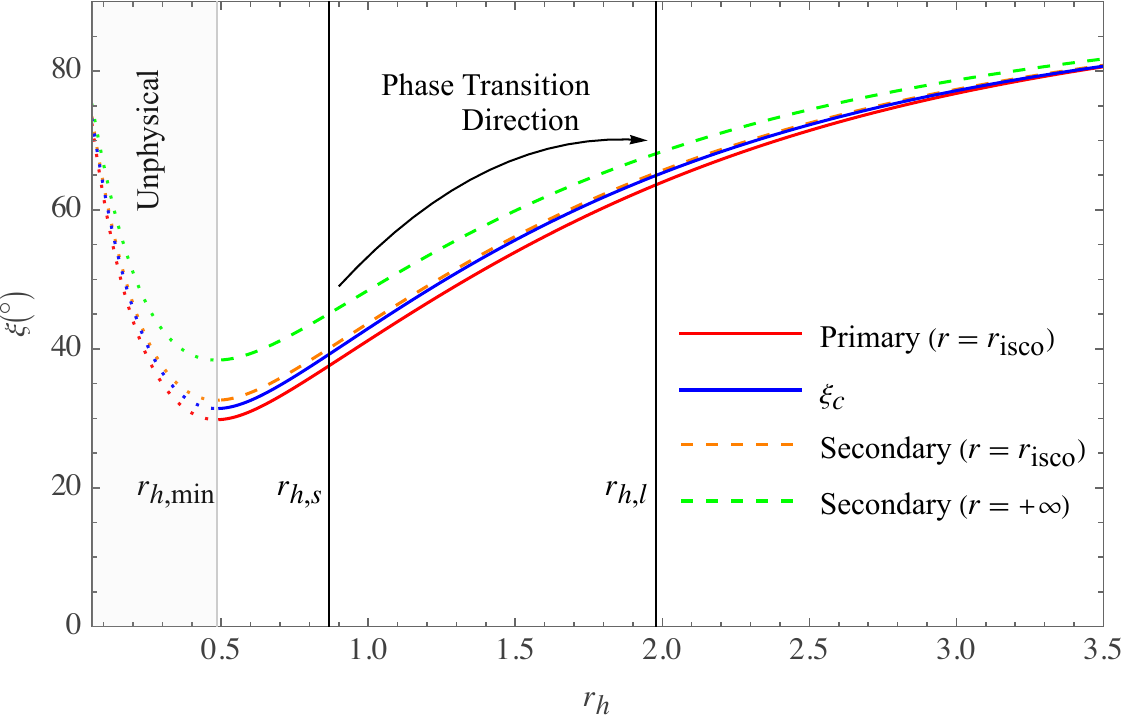}

 \caption{Evolution of the black hole shadow angle $\xi_{\rm c}$ and the accretion-disk image angle $\xi_{\rm n}$
 as functions of the horizon radius $r_h$ in the isobaric process, with $Q=0.5$, $P=0.8\,P_{c}$, and $r_0=50$. The horizontal axis $r_h$, the quantities $r_{h,s}$, $r_{h,l}$, and $r_{h,\rm min}$, as well as the fixed parameters $Q$ and $r_0$, all have length dimension $L$, while the fixed parameter $P$ has dimension $L^{-2}$ and the vertical axis $\xi$ is dimensionless.}
 \label{fig:dyy}
\end{figure}

\begin{figure}[htbp]
 \centering
 \includegraphics[width=1\linewidth]{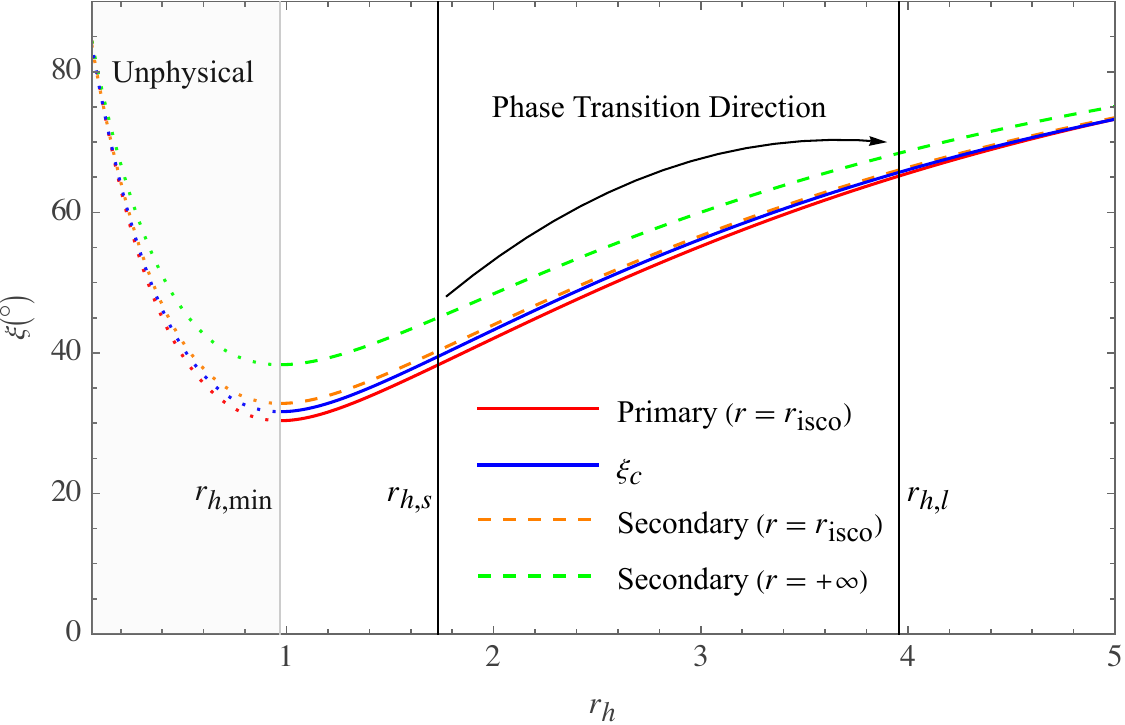}

 \caption{Evolution of the black hole shadow angle $\xi_{\rm c}$ and the accretion-disk image angle $\xi_{\rm n}$
 as functions of the horizon radius $r_h$ in the isobaric process, with $Q=1$, $P=0.8\,P_{c}$, and $r_0=50$. The horizontal axis $r_h$, the quantities $r_{h,s}$, $r_{h,l}$, and $r_{h,\rm min}$, as well as the fixed parameters $Q$ and $r_0$, all have length dimension $L$, while the fixed parameter $P$ has dimension $L^{-2}$ and the vertical axis $\xi$ is dimensionless.}
 \label{fig:dy}
\end{figure}

\begin{figure}[htbp]
 \centering
 \includegraphics[width=1\linewidth]{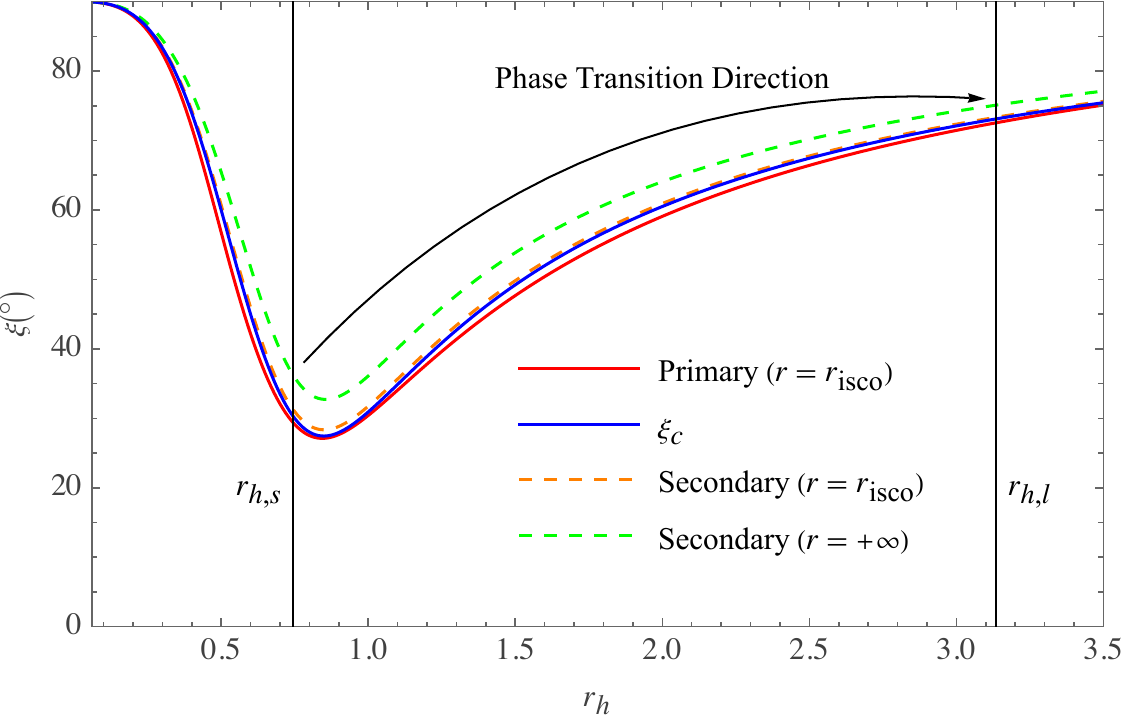}

 \caption{Evolution of the black hole shadow angle $\xi_{\rm c}$ and the accretion-disk image angle $\xi_{\rm n}$
 as functions of the horizon radius $r_h$ in the isothermal process, with $Q=0.5$, $T_0=0.8\,T_{c}$, and $r_0=50$. The horizontal axis $r_h$, the quantities $r_{h,s}$ and $r_{h,l}$, as well as the fixed parameters $Q$ and $r_0$, all have length dimension $L$, while the fixed parameter $T_0$ has dimension $L^{-1}$ and the vertical axis $\xi$ is dimensionless.}
 \label{fig:dww}
\end{figure}

\begin{figure}[htbp]
 \centering
 \includegraphics[width=1\linewidth]{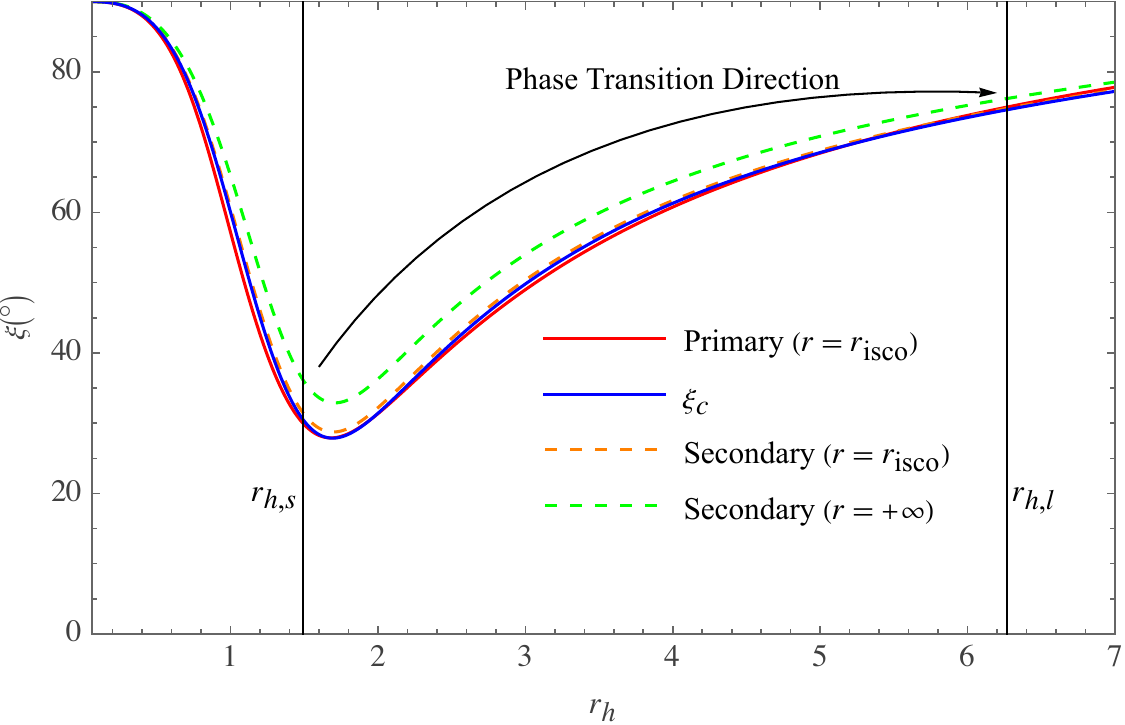}

 \caption{Evolution of the black hole shadow angle $\xi_{\rm c}$ and the accretion-disk image angle $\xi_{\rm n}$
 as functions of the horizon radius $r_h$ in the isothermal process, with $Q=1$, $T_0=0.8\,T_{c}$, and $r_0=50$. The horizontal axis $r_h$, the quantities $r_{h,s}$ and $r_{h,l}$, as well as the fixed parameters $Q$ and $r_0$, all have length dimension $L$, while the fixed parameter $T_0$ has dimension $L^{-1}$ and the vertical axis $\xi$ is dimensionless.}
 \label{fig:dw}
\end{figure}

Our numerical results are shown in Figs.~\ref{fig:dyy}--\ref{fig:dw}, with Figs.~\ref{fig:dyy} and \ref{fig:dy} corresponding to the isobaric process and Figs.~\ref{fig:dww} and \ref{fig:dw} to the isothermal process. In all panels, the shadow angle $\xi_{\rm c}(r_h)$ is plotted as a blue solid curve. For convenience in the subsequent analysis of phase transitions, we fix the pressure and temperature to $P=0.8\,P_c<P_c$ and $T=0.8\,T_c<T_c$, where $P_c$ and $T_c$ denote the critical pressure and temperature. Under this setup, the system will undergo a phase transition. These figures clearly show the following behavior: In the isobaric process, once the nonphysical branch is truncated at $r_{h,{\rm min}}$, $\xi_{\rm c}(r_h)$ increases strictly monotonically and approaches $\xi_{\rm c}=90^\circ$ asymptotically. By contrast, in the isothermal process $\xi_{\rm c}(r_h)$ exhibits a nonmonotonic behavior, decreasing at small $r_h$ and increasing at larger $r_h$, while approaching $90^\circ$ in both limits. These numerical results are in precise agreement with Eq.~\eqref{eq:hg2} and the above analytic arguments.

In the same figures we also display the numerical results for the angular sizes $\xi_n(r_h)$ of accretion-disk images. The red solid curves correspond to the primary images produced by emitters on the $r=r_{\rm isco}$ orbit, while the orange and green dashed curves represent the secondary images generated by emitters on the $r=r_{\rm isco}$ and $r\!\to\!+\infty$ orbits, respectively. Note that we do not plot the primary image associated with $r\!\to\!+\infty$, since it appears excessively large to the observer, with an angular size that never falls below $90^\circ$.

These plots reveal a striking feature: the monotonicity of $\xi_n(r_h)$ is identical to that of the shadow angle $\xi_{\rm c}(r_h)$, and their extrema occur at the same values of $r_h$. This demonstrates that the nonmonotonic evolution of accretion-disk images can serve as the same diagnostic for distinguishing the isobaric and isothermal processes as the black hole shadow. From an observational perspective, accretion disks are particularly appealing because they are intrinsically luminous and do not require an idealized background light source, making them a more accessible probe of black hole thermodynamic evolution.

We also note an intriguing phenomenon present in these figures: the angular size of the primary image associated with $r=r_{\rm isco}$ can be smaller than that of the corresponding secondary image. Similar effects have been discussed in previous studies, and we defer a detailed investigation of this behavior to future work.

Note that the above discussion has not yet taken black hole phase transitions into account, whose presence breaks the otherwise continuous evolution of black hole images and may therefore encode additional thermodynamic information. To this end, the small and large black hole radii at the phase transition, $r_{h,s}$ and $r_{h,l}$, determined via the Maxwell equal-area construction, are marked in Figs.~\ref{fig:dyy}-\ref{fig:dw}\footnote{Note that this does not mean that time itself also undergoes a sudden jump. It only means that the time reading jumps from $t(r_{h,s})$ to $t(r_{h,l})$.}. At these radii the otherwise continuous evolution of the image size $\xi(r_h)$ (including both $\xi_{\rm c}(r_h)$ and $\xi_n(r_h)$) is interrupted by a discontinuous jump, which immediately raises two natural questions. First, does the black hole image always sudden increase across the phase transition, namely $\xi(r_{h,l})>\xi(r_{h,s})$? Second, is the monotonicity of $\xi(r_h)$ the same before and after the phase transition?

For the isobaric process the answers are straightforward. Since $\xi(r_h)$ is strictly monotonically increasing in the isobaric process, one necessarily has $\xi(r_{h,l})>\xi(r_{h,s})$, implying a sudden increase of the image size across the phase transition and supporting earlier proposals that black hole phase transitions may be diagnosed through sudden changes in optical observables. Moreover, $\xi(r_h)$ remains monotonically increasing on both sides of the phase transition.

In the isothermal process, the behavior of $\xi(r_h)$ differs significantly due to its nonmonotonic nature. In this case, several distinct scenarios emerge based on the relative ordering of the small black hole radius $r_{h,s}$, the large black hole radius $r_{h,l}$, and the extremum position $r_{h,e}$ of $\xi(r_h)$ as follows:
\begin{enumerate}[(i)]
\item if $r_{h,s}<r_{h,l}<r_{h,e}$, then $\xi(r_{h,l})<\xi(r_{h,s})$, with $\xi(r_h)$ decreasing before the phase transition and exhibiting a decrease followed by an increase afterward;
\item if $r_{h,s}\le r_{h,e}\le r_{h,l}$, then $\xi(r_{h,l})$ may be larger than, smaller than, or equal to $\xi(r_{h,s})$, while $\xi(r_h)$ decreases before the phase transition and increases afterward;
\item if $r_{h,e}<r_{h,s}<r_{h,l}$, then $\xi(r_{h,l})>\xi(r_{h,s})$, with $\xi(r_h)$ exhibiting a decrease followed by an increase before the phase transition and a monotonic increase afterward.
\end{enumerate}

The configurations displayed in Figs.~\ref{fig:dww} and \ref{fig:dw} correspond to case (ii) with $\xi(r_{h,l})>\xi(r_{h,s})$. It is worth noting that all three cases preserve the nonmonotonic evolution of black hole images in the isothermal process, indicating that the inclusion of the phase transition does not invalidate the criterion based on nonmonotonicity for distinguishing the isobaric from the isothermal processes. To clarify which of these possibilities can actually occur, we proceed in two steps: we first analyze the relative ordering of $r_{h,e}$, $r_{h,s}$, and $r_{h,l}$, and then examine the ordering between $\xi(r_{h,l})$ and $\xi(r_{h,s})$.

To implement these two steps, the most straightforward strategy is to determine $r_{h,e}$, $r_{h,s}$, $r_{h,l}$ and the corresponding image sizes $\xi(r_{h,s})$ and $\xi(r_{h,l})$ for all admissible black hole parameters. For RN--AdS black holes in the isothermal process, the admissible parameters are the charge $Q$ and the temperature $T_{0}$. Fortunately, the dependence on the charge $Q$ can be eliminated by introducing reduced (dimensionless) variables,
\begin{equation}
 \tilde r_{h}=\frac{r_{h}}{r_{h,c}},\quad
 \tilde P=\frac{P}{P_{c}},\quad
 \tilde T=\frac{T}{T_{c}},\quad
 \tilde M=\frac{M}{M_{c}},\quad
\end{equation}
where $r_{h,c}$, $P_c$, and $T_c$ are the critical horizon radius, pressure, and temperature given in Eqs.~(\ref{eq:rhp}) and~(\ref{eq:rht}). The critical mass $M_c$ follows from Eq.~\eqref{eq:mrh} by substituting the values of $r_{h,c}$ and $T_c$, yielding $M_c=2\sqrt{6}\,Q/3$. In addition, since the observer position $r_{0}$ is freely chosen and does not possess an intrinsic critical value, we rescale it by the critical horizon radius,
\begin{equation}
 \tilde r_{0}=\frac{r_{0}}{r_{h,c}}.
\end{equation}
Substituting these reduced variables into Eqs.~\eqref{eq:mrh} and \eqref{eq:prh} yields
\begin{align}
 \tilde M(\tilde r_{h}) &= \frac{1 + 3\,\tilde r_{h}^{2} + 2\,\tilde T_{0}\tilde r_{h}^{3}}{6\,\tilde r_{h}},
 \\[4pt]
 \tilde P(\tilde r_{h}) &= \frac{1 - 6\,\tilde r_{h}^{2} + 8\,\tilde T_{0}\tilde r_{h}^{3}}{3\,\tilde r_{h}^{4}},
 \label{eq:reprh}
\end{align}
which are manifestly independent of the charge $Q$, implying that all results obtained from them are universal and valid for arbitrary $Q$. Likewise, the shadow angle $\xi_{\rm c}$ can be expressed in terms of the reduced variables and thus no longer depends explicitly on the charge $Q$,
\begin{equation}
 \xi_{\rm c} =\xi_{\rm c}\!\left(\tilde r_{h};\,\tilde r_{0},\,\tilde T_{0}\right).
 \label{eq:rexic}
\end{equation}

After this reduction, only a single control parameter, the reduced temperature $\tilde T_{0}$ (equivalently $T_{0}$), remains, allowing all relevant cases to be systematically explored through a numerical scan over $\tilde T_{0}$. It should be noted that the requirement $\tilde P(\tilde r_{h}) \ge 0$ for all admissible $\tilde r_{h}$ imposes a lower bound on the temperature, $\tilde T_{0}\ge \sqrt{2}/2$, so that the physically relevant range is $\tilde T_{0}\in[\sqrt{2}/2,\,1]$.

\begin{figure}[htbp]
 \centering
 \includegraphics[width=1\linewidth]{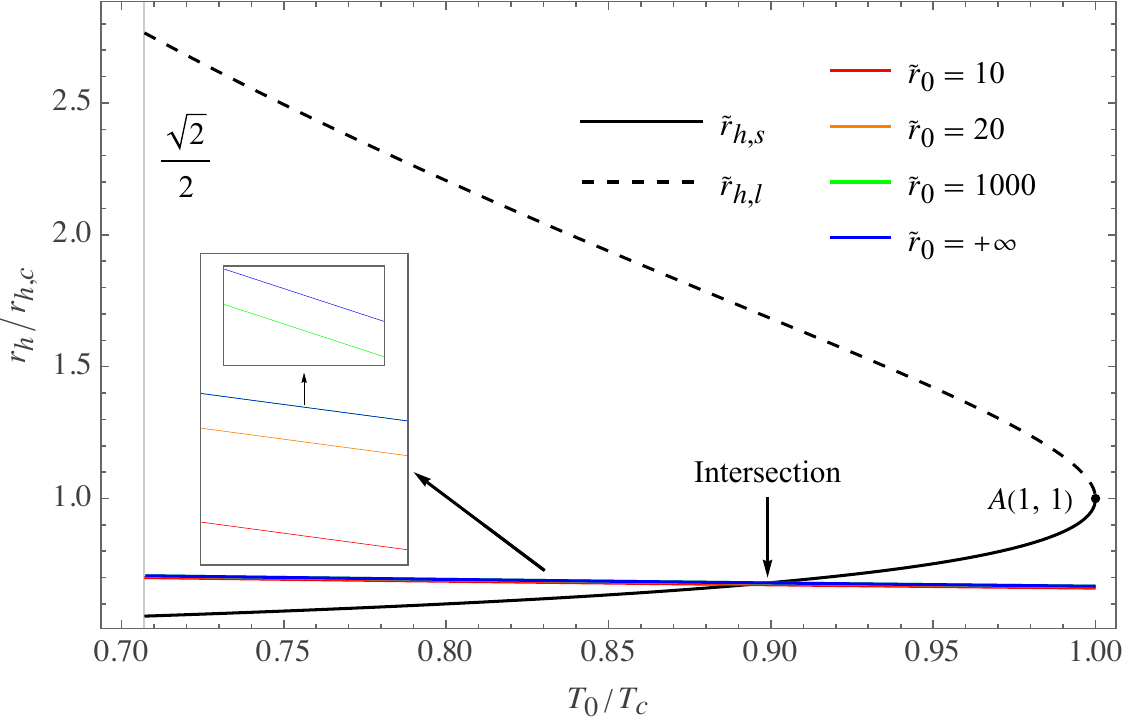}

 \caption{Reduced small, large, and extremal radius curves, $\tilde r_{h,s}$, $\tilde r_{h,l}$, and $\tilde r_{h,e}$, as functions of the reduced temperature $\tilde T_{0}$. All quantities are dimensionless.}
 \label{fig:gx}
\end{figure}

We now address the first step outlined above. Applying the Maxwell equal-area construction to Eq.~\eqref{eq:reprh} yields the reduced small and large black hole radii $\tilde r_{h,s}(\tilde T_{0})$ and $\tilde r_{h,l}(\tilde T_{0})$, while solving $\partial\xi_{\rm c}/\partial\tilde r_{h}=0$ from Eq.~\eqref{eq:rexic} determines the extremal radius $\tilde r_{h,e}(\tilde r_{0},\tilde T_{0})$. These quantities are displayed in Figure~\ref{fig:gx}, where the black solid and dashed curves denote $\tilde r_{h,s}(\tilde T_{0})$ and $\tilde r_{h,l}(\tilde T_{0})$, respectively, meeting at point $A(1,1)$ that corresponds to $r_{h,s}=r_{h,l}=r_{h,c}$. The four colored solid curves correspond to $\tilde r_{h,e}(\tilde r_{0},\tilde T_{0})$ for different values of $\tilde r_{0}$.

\begin{figure}[htbp]
 \centering
 \includegraphics[width=1\linewidth]{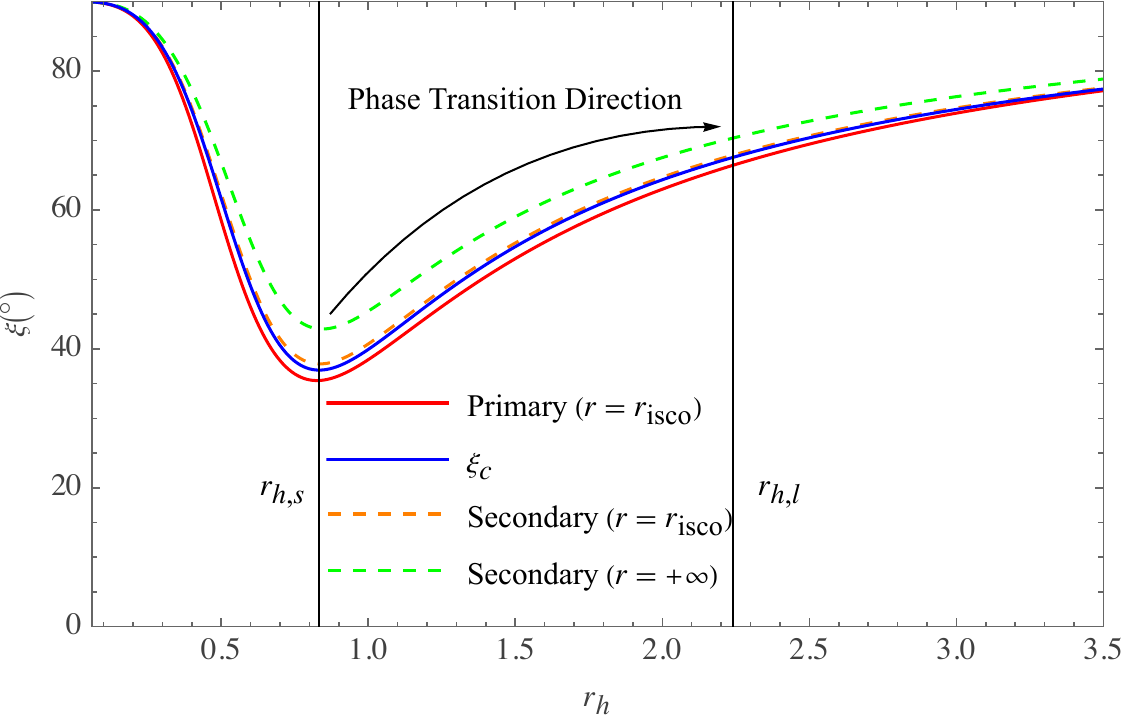}

 \caption{Evolution of the black hole shadow angle $\xi_{\rm c}$ and the accretion-disk image angle $\xi_{\rm n}$
 as functions of the horizon radius $r_h$ in the isothermal process, with $Q=0.5$, $T_0 \approx 0.89251\,T_c$ ($\tilde T_{0}=\tilde T_{0,c}$), and $r_0=50$. The horizontal axis $r_h$, the quantities $r_{h,s}$ and $r_{h,l}$, as well as the fixed parameters $Q$ and $r_0$, all have length dimension $L$, while the fixed parameter $T_0$ has dimension $L^{-1}$ and the vertical axis $\xi$ is dimensionless.}
 \label{fig:zj}
\end{figure}

\begin{figure}[htbp]
 \centering
 \includegraphics[width=1\linewidth]{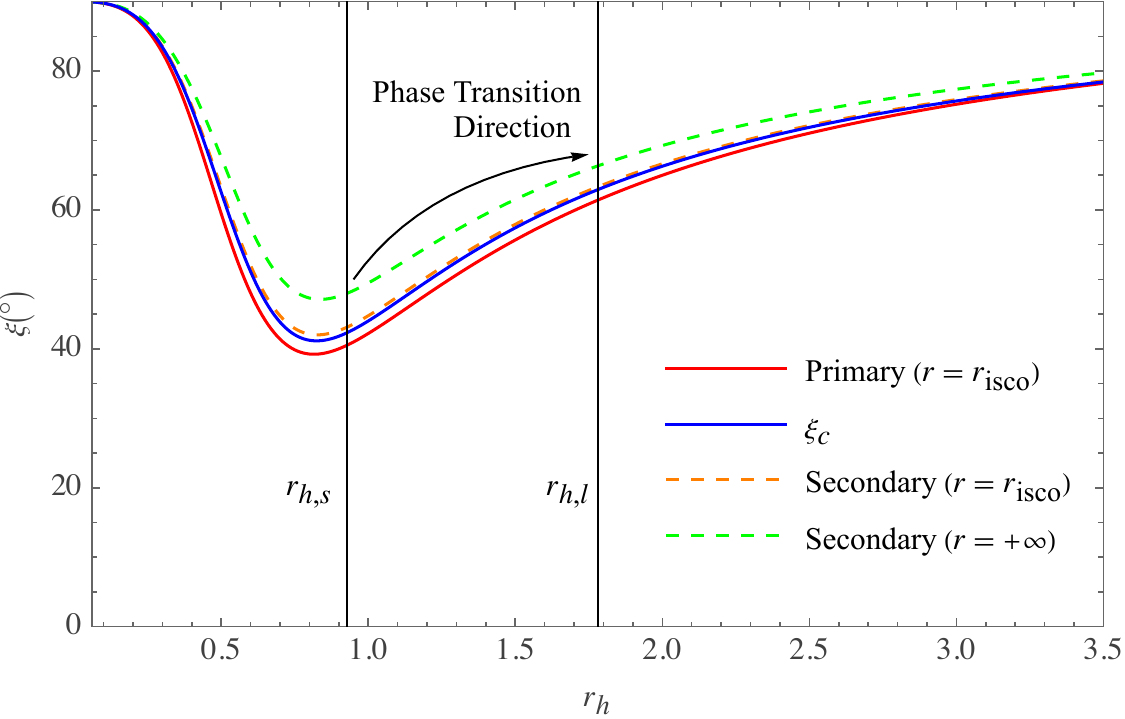}

 \caption{Evolution of the black hole shadow angle $\xi_{\rm c}$ and the accretion-disk image angle $\xi_{\rm n}$
 as functions of the horizon radius $r_h$ in the isothermal process, with $Q=0.5$, $T_0=0.95\,T_c$ ($\tilde T_{0}>\tilde T_{0,c}$), and $r_0=50$. The horizontal axis $r_h$, the quantities $r_{h,s}$ and $r_{h,l}$, as well as the fixed parameters $Q$ and $r_0$, all have length dimension $L$, while the fixed parameter $T_0$ has dimension $L^{-1}$ and the vertical axis $\xi$ is dimensionless.}
 \label{fig:dwy}
\end{figure}

It is evident from Fig.~\ref{fig:gx} that throughout the allowed range of $\tilde T_{0}$ one always has $\tilde r_{h,l}>\tilde r_{h,e}$, so that the previously discussed case~(i) never occurs. Cases~(ii) and~(iii) are separated by the intersection of $\tilde r_{h,s}$ and $\tilde r_{h,e}$ shown in the figure. We define the temperature at this intersection as the critical reduced temperature $\tilde T_{0,c}$, which gives rise to the following two regimes:
\begin{enumerate}[(a)]
\item For $\tilde T_{0}\le \tilde T_{0,c}$ one has $\tilde r_{h,s}\le \tilde r_{h,e}<\tilde r_{h,l}$, corresponding to case~(ii), in which the shadow angle $\xi_{\rm c}(r_h)$ decreases monotonically before the phase transition and increases monotonically afterward;
\item For $\tilde T_{0}>\tilde T_{0,c}$ one finds $\tilde r_{h,e}<\tilde r_{h,s}<\tilde r_{h,l}$, corresponding to case~(iii), where $\xi_{\rm c}(r_h)$ first decreases and then increases before the phase transition and increases monotonically afterward.
\end{enumerate}
These results demonstrate that black hole phase transition not only induces a discontinuous jump in the image size, but also, under the isothermal process, allows one to distinguish two distinct temperature regimes, $\tilde T_{0}\le \tilde T_{0,c}$ and $\tilde T_{0}>\tilde T_{0,c}$, from the black hole shadow size, thereby revealing additional thermodynamic information encoded in black hole images. Moreover, the same conclusion is expected to hold for accretion-disk images, since their angular sizes $\xi_{n}(r_h)$ exhibit the same qualitative behavior as $\xi_{\rm c}(r_h)$. To verify this, we numerically plot $\xi(r_h)$ for $\tilde T_{0}=\tilde T_{0,c}$ (Fig.~\ref{fig:zj}) and for $\tilde T_{0}>\tilde T_{0,c}$ (Fig.~\ref{fig:dwy}), with results fully consistent with our theoretical expectation.

\begin{figure}[htbp]
 \centering
 \includegraphics[width=1\linewidth]{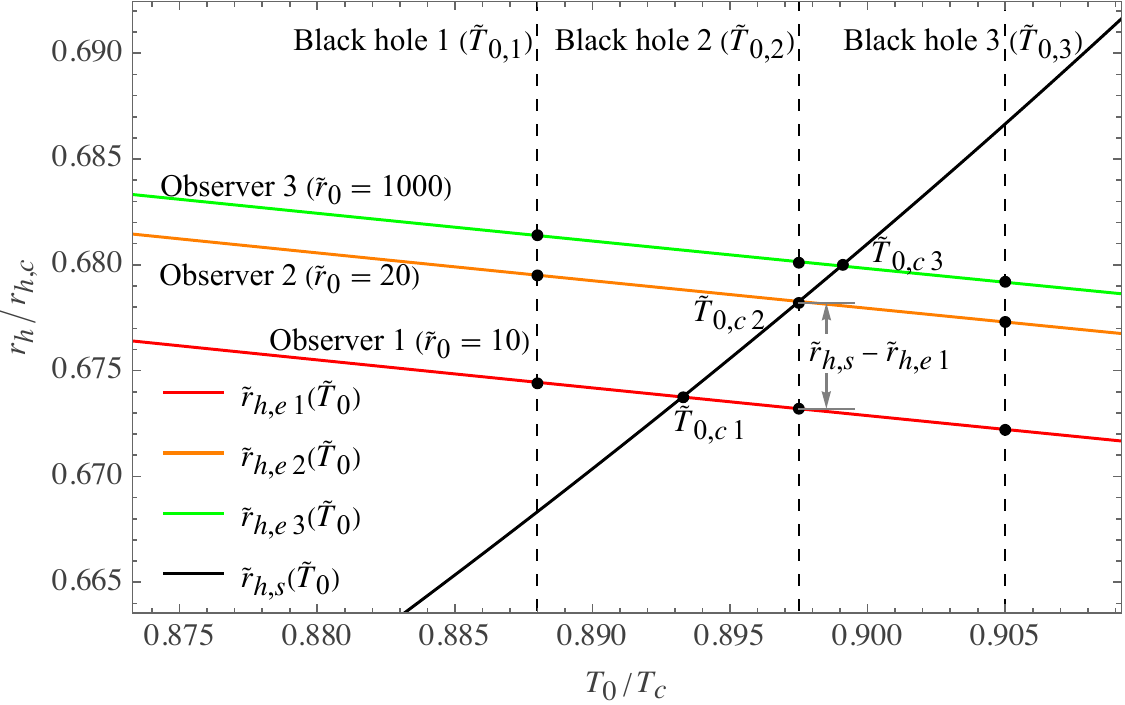}

 \caption{Schematic illustration of three observers at different positions observing black holes at three different temperatures. All quantities are dimensionless.}
 \label{fig:new1}
\end{figure}

It is crucial to distinguish the critical reduced temperature from the intrinsic reduced temperature of the black hole. As shown by the four colored curves in Fig.~\ref{fig:gx}, the critical reduced temperature depends on the observer position $r_0$, as indicated by the intersections of the colored curves with the black solid curve, although this dependence is weak. In contrast, the black hole's intrinsic reduced temperature, determined by the ratio $T_0/T_c$, is independent of $r_0$. This distinction gives rise to the following interesting phenomenon. For clarity, we enlarge the region around the points labeled ``intersections'' in Fig.~\ref{fig:gx} and present it as Fig.~\ref{fig:new1}. The three colored curves represent $\tilde r_{h,e}(\tilde T_{0})$ for observers at three different positions, while the three vertical dashed lines correspond to three black holes at different temperatures. The black solid curve denotes $\tilde r_{h,s}(\tilde T_{0})$, and its intersections with the three colored curves determine the three corresponding critical reduced temperatures: $\tilde T_{0,c1}$, $\tilde T_{0,c2}$, and $\tilde T_{0,c3}$. For black hole $1$, it is clear that observers at all three positions would conclude that the black hole is in the regime $\tilde T_{0,1}\le \tilde T_{0,c}$ (including $\tilde T_{0,c1}$, $\tilde T_{0,c2}$, and $\tilde T_{0,c3}$), namely case~(ii). For black hole $3$, all three observers would conclude that the black hole is in the regime $\tilde T_{0,3}>\tilde T_{0,c}$, namely case~(iii). For black hole $2$, however, observers $2$ and $3$ would identify it as case~(ii), whereas observer $1$ would identify it as case~(iii). This shows that, for the same black hole, if its reduced temperature happens to coincide with the critical reduced temperature for an observer at a certain position, different observers may see different black hole image evolution patterns. It should be noted, however, that the difference $\tilde r_{h,s}-\tilde r_{h,e1}$ corresponding to observer $1$ is theoretically very small. This means that the rising segment to the left of the vertical line $r_{h,s}$ in Fig.~\ref{fig:dwy} is very short. As a result, observer $1$ may require an extremely high-precision detector to observe the brief increase in the black hole image size. Therefore, from the perspective of astronomical observation, it is safer to deploy multiple high-precision detectors at different positions in order to determine the black hole temperature more reliably.

\begin{figure}[htbp]
 \centering
 \includegraphics[width=1\linewidth]{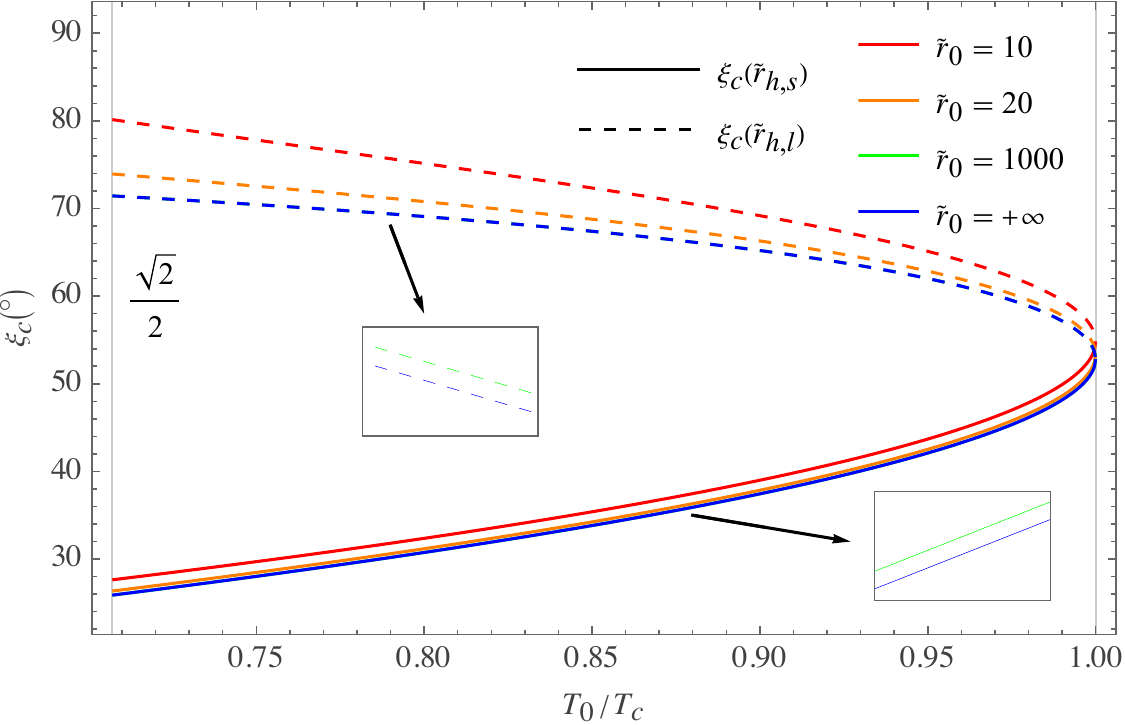}

 \caption{Shadow angle $\xi_{\rm c}$ evaluated at the reduced small and large black hole radii $\tilde r_{h,s}$ and $\tilde r_{h,l}$ as a function of the reduced temperature $\tilde T_{0}$. All quantities are dimensionless.}
 \label{fig:xict}
\end{figure}

We now turn to the second step outlined above. This is straightforward: inserting the previously obtained reduced radii $\tilde r_{h,s}(\tilde T_{0})$ and $\tilde r_{h,l}(\tilde T_{0})$ into Eq.~\eqref{eq:rexic} directly yields the corresponding shadow angles. The results are displayed in Fig.~\ref{fig:xict}, where the colored solid and dashed curves represent $\xi_{\rm c}\!\left(\tilde r_{h,s}\right)$ and $\xi_{\rm c}\!\left(\tilde r_{h,l}\right)$, respectively. One clearly finds that, for both $\tilde T_{0}\le \tilde T_{0,c}$ [case~(ii)] and $\tilde T_{0}>\tilde T_{0,c}$ [case~(iii)],
\begin{equation}
 \xi_{\rm c}\!\left(\tilde r_{h,l}\right)\ge
 \xi_{\rm c}\!\left(\tilde r_{h,s}\right),
 \label{eq:xbty}
\end{equation}
with equality attained only at $\tilde T_{0}=1$, where $r_{h,s}=r_{h,l}=r_{h,c}$. Therefore, although $\xi_{\rm c}(r_h)$ exhibits nonmonotonic behavior during the isothermal process, the black hole shadow always undergoes a sudden increase across the phase transition, similar to the isobaric case. While this holds for observers at any position, the jump in image size is larger for those nearer to the black hole. Interestingly, this conclusion also implies that sudden increases of the shadow alone can signal the occurrence of the phase transition, but are insufficient to distinguish between isobaric and isothermal processes. The same conclusion is expected to hold for accretion-disk images.

\section{Conclusions}
\label{sec4}

In this work we have investigated how the images of RN--AdS black holes evolve along isobaric and isothermal thermodynamic processes. Our results show that black hole images encode not only information about the phase transition, but also about the underlying thermodynamic process, and even about the temperature in the isothermal case.

Phase transition information encoded in black hole images has long been recognized, particularly in isobaric process, where numerous studies have shown that a thermodynamic phase transition manifests itself as a sudden change in the image size. In this paper, we extend this conclusion to the isothermal process and demonstrate that, in both thermodynamic processes, the phase transition always leads to a sudden increase in the black hole image, rather than a decrease or no change. While this result further confirms the robustness of using image discontinuities to detect black hole phase transitions, it also implies that such sudden changes alone are generally insufficient to distinguish between the underlying isobaric and isothermal processes.

Fortunately, we find that black hole images exhibit qualitatively different evolution patterns in the two thermodynamic processes. Along isobars, the image size grows monotonically with the horizon radius $r_h$, whereas along isotherms it first decreases and then increases. This contrast implies that the nonmonotonic behavior of the image size itself provides a diagnostic for distinguishing isobaric from isothermal processes.

Further, we find that when the black hole evolves along the isothermal path, the nonmonotonic behavior of the image size leads to an extremal radius. Depending on its ordering relative to the small and large black hole radii at the phase transition, three distinct evolutionary patterns are possible in principle. We demonstrate, however, that only two of them are physically realized, and that they are separated by a critical reduced temperature $\tilde T_{0,c}$. For $\tilde T_{0}\le \tilde T_{0,c}$, the image size decreases monotonically with $r_h$ before the phase transition and increases monotonically afterward. By contrast, for $\tilde T_{0}>\tilde T_{0,c}$, the image size first decreases and then increases with $r_h$ before the phase transition, while still increasing monotonically afterward. These two clearly distinct behaviors provide a direct observational criterion for discriminating between different temperature ranges in the isothermal process. It should be noted, however, that these two different behaviors may be observed simultaneously by observers at different positions. If this occurs, it means that the reduced temperature of the black hole happens to coincide with the critical reduced temperature for one of the observers. Therefore, from the perspective of astronomical observation, it is preferable to place multiple high-precision detectors at different distances from the black hole in order to obtain its temperature information more reliably.

Our results further enrich the thermodynamic information extractable from black hole images, and suggest new avenues for testing black hole thermodynamics observationally. Moreover, it is worth emphasizing that we expect the framework adopted here, which combines phase transitions with the nonmonotonic evolution of black hole images, to be applicable to other black hole models and may uncover additional thermodynamic information encoded in black hole images.

%-------------------------------------------------------------
\begin{acknowledgments}
We would like to thank Zhongzhinan Dong and Jiawei Chen for helpful discussions. This work is supported in part by NSFC Grants No. 12165005 and No. 11961131013.
\end{acknowledgments}
%-------------------------------------------------------------

\appendix

\section{Relation between photon emission angle and impact parameter}
\label{appa}

In this appendix, we derive Eq.~\eqref{eq:sinxi} and provide an intuitive interpretation of the photon emission angle $\xi$. We place a static observer at
\begin{equation}
 \left(t,r,\theta,\phi\right)
 = \left(t_0, r_0, \theta_0, \phi_0\right),
\end{equation}
whose four-velocity $U^a$ is parallel to the timelike Killing vector $(\partial_t)^a$. Here and in what follows, Latin letters $a,b,c,\ldots$ denote abstract indices, while Greek letters $\mu,\nu,\ldots$ denote component indices (or concrete indices), following Refs.~\cite{Wald:1984rg,Liang:2023ahd}. Taking into account the normalization condition for a timelike four-velocity, $U_a U^a = -1$, the observer's four-velocity can be written as
\begin{equation}
 U^a=\frac{1}{\sqrt{f(r_0)}}\, (\partial_t)^a.
\end{equation}
A natural orthonormal tetrad $\{(e_{\mu})^a\}$ adapted to this observer is
\begin{align}
 (e_{t})^a
 &= \frac{1}{\sqrt{f(r_0)}}\, (\partial_t)^a, \\
 (e_{r})^a
 &= \sqrt{f(r_0)}\, (\partial_r)^a, \\
 (e_{\theta})^a
 &= \frac{1}{r_0}\, (\partial_{\theta})^a, \\
 (e_{\phi})^a
 &= \frac{1}{r_0\sin\theta_0}\, (\partial_{\phi})^a.
\end{align}
The tetrad $\{(e_{\mu})^a\}$ provides the local inertial frame in which all local measurements are defined.

For a photon at the observer's position, its four-momentum $K^a$ can be conveniently decomposed in the tetrad $\{(e_{\mu})^a\}$ as
\begin{equation}
 K^a=\omega (e_{t})^a+k^i (e_{i})^a\equiv\omega (e_{t})^a+k^a,
\end{equation}
where $i$ runs over $r$, $\theta$, and $\phi$, and the quantities $\omega$ and $k^a$ correspond to the photon energy and three-momentum measured by the observer, respectively. They are obtained by
\begin{equation}
 \omega = -\eta_{ab} K^a U^b = -K_b (e_t)^b,
\end{equation}
and
\begin{equation}
 k^i = k_i = \eta_{ab} K^a (e_{i})^b = K_b (e_{i})^b,
\end{equation}
where $\eta_{ab}$ denotes the Minkowski metric. Note that the spacetime metric is originally $g_{ab}$, but in the orthonormal tetrad $\{(e_{\mu})^a\}$, it reduces to $\eta_{ab}$.

\begin{figure}[htbp]
 \centering
 \includegraphics[width=1\linewidth]{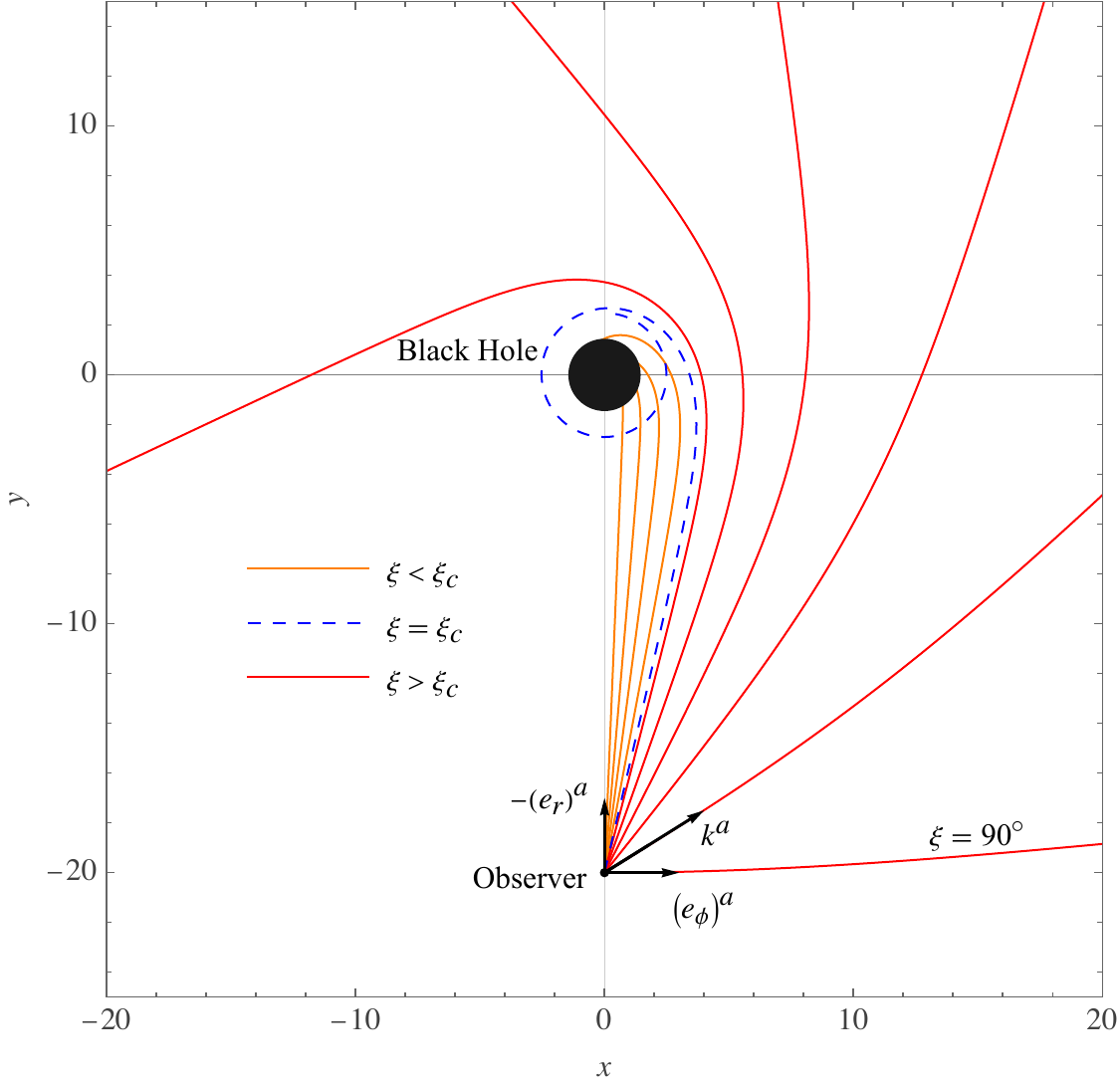}

 \caption{Trajectories of photons emitted from the observer at different angles on the equatorial plane in the RN--AdS spacetime. The horizontal axis $x$ and the vertical axis $y$ have length dimension $L$, while $\xi$ and $\xi_c$ are dimensionless.}
 \label{fig:xy}
\end{figure}

We now introduce the angle $\xi$ to characterize the direction of the photon in the observer's sky. It is defined as the angle between the tetrad vector $-(e_{r})^a$ and the photon three-momentum $k^a$ (see Fig.~\ref{fig:xy}). Since we consider a spherically symmetric spacetime, we restrict the photon motion to the equatorial plane $(\theta=\pi/2)$ for later convenience without loss of generality. Under this restriction, $k^\theta=0$, and hence the photon three-momentum $k^a$ can be written as
\begin{equation}
 k^a = k^r (e_r)^a + k^\phi (e_\phi)^a.
\end{equation}
The angle $\xi$ can then be conveniently expressed as
\begin{equation}
 \sin\xi
 =\frac{{\rm Abs}(k^{\phi})}{|k|},
 \label{eq:a1}
\end{equation}
where ${\rm Abs}(k^{\phi})$ represents the absolute value of $k^{\phi}$ and $|k|$ denotes the length of $k^a$ measured by $\eta_{ab}$. Since $K^a$ is a null vector, we have
\begin{equation}
 0 = \eta_{ab} K^a K^b
 = -\omega^2 + |k|^2.
\end{equation}
Therefore, $|k|=\omega$. Meanwhile, noting that the spacetime admits two Killing vector fields, $(\partial_t)^a$ and $(\partial_{\phi})^a$, we have
\begin{equation}
 \omega = -K_b (e_t)^b
 = -\frac{1}{\sqrt{f(r_0)}}\, K_b (\partial_t)^b
 = \frac{E}{\sqrt{f(r_0)}},
\end{equation}
and
\begin{equation}
 k^{\phi}=K_b (e_{\phi})^b
 = \frac{1}{r_0\sin(\pi/2)}\,K_b(\partial_{\phi})^b
 = \frac{L}{r_0},
\end{equation}
where $E$ and $L$ are the conserved energy and angular momentum of the photon associated with the Killing vector fields $(\partial_t)^a$ and $(\partial_{\phi})^a$, respectively. Finally, using the relations above, Eq.~\eqref{eq:a1} can be rewritten as
\begin{equation}
 \sin\xi
 = \frac{L\,\sqrt{f(r_0)}}{E\,r_0}
 = \frac{b\,\sqrt{f(r_0)}}{r_0},
 \label{eq:a2}
\end{equation}
which is exactly Eq.~\eqref{eq:sinxi}. Note that here we always take the photon to move in the direction of increasing $\phi$, and therefore $L>0$. When the photon impact parameter $b$ is exactly equal to the critical impact parameter $b_{\rm c} = {r_{\rm ps}}/{\sqrt{f(r_{\rm ps})}}$, the corresponding angle becomes the critical angle $\xi_{\rm c}$. This corresponds to Eq.~\eqref{eq:xic},
\begin{equation}
 \xi_{\rm c} = \arcsin\!\left(\frac{b_{\rm c}\,\sqrt{f(r_0)}}{r_0}\right).
\end{equation}
In Fig.~\ref{fig:xy}, we plot the trajectories of photons emitted from the observer at different angles. It is easy to see that these trajectories exhibit the following typical features: For $\xi<\xi_{\rm c}$ (or equivalently $b<b_{\rm c}$), the photon falls into the black hole. For $\xi=\xi_{\rm c}$ ($b=b_{\rm c}$), the photon undergoes an asymptotically circular motion around the black hole. For $\xi>\xi_{\rm c}$ ($b>b_{\rm c}$), the photon turns around at the periastron and then escapes outward. This is qualitatively consistent with the case of an observer at infinity in the Schwarzschild spacetime. It is worth noting that one can readily see from Eq.~\eqref{eq:a2} that the photon with $\xi=90^\circ$ has the maximum impact parameter
\begin{equation}
 b_{\max}
 = \frac{r_{0}}{\sqrt{f(r_{0})}}
 = \frac{1}{\sqrt{U(r_{0})}}.
\end{equation}
It satisfies $b_{\max}< \ell$ for $r_0 \in (r_{\rm ps},+\infty)$. Therefore, in the RN--AdS spacetime, photons with $b \geq \ell$ do not contribute to the black hole image.

%%=============================================================================================

%%=============================================================================================

\end{document}